\documentclass[useAMS,usedcolumn]{mn2e}
\usepackage{supertabular,epsf,rotate}
\newcommand{\f}{\phantom{2}}
\newcommand{\mc}{\multicolumn}

\begin{document}

\title[The $K-z$ relation for radio galaxies in the 7CRS]
{Near-infrared imaging and the $K-z$ relation for radio galaxies in
the 7C Redshift Survey}

\author[Willott et al.]
{Chris J.\ Willott$^{1,2}$\footnotemark, 
Steve Rawlings$^{2}$, 
Matt J.\,Jarvis$^{3}$, 
Katherine M.\ Blundell$^{2}$\\
$^{1}$Herzberg Institute of Astrophysics, National Research Council, 5071 West Saanich Rd, Victoria, B.C. V9E 2E7, Canada\\
$^{2}$Astrophysics, Department of Physics, Keble Road, Oxford, OX1
3RH, U.K. \\
$^{3}$Sterrewacht Leiden, Postbus 9513, 2300 RA Leiden, the
Netherlands \\}

\maketitle

\begin{abstract} 

We present $K$-band imaging of all 49 radio galaxies in the 7C--I and
7C--II regions of the 7C Redshift Survey (7CRS). The low--frequency
(151 MHz) selected 7CRS sample contains all sources with
flux-densities $S_{151} > 0.5$ Jy in three regions of the sky.  We
combine the $K$-band magnitudes of the 7CRS radio galaxies with those
from the 3CRR, 6CE and 6C$^\star$ samples to investigate the nature of
the relationship between $K$-magnitude and redshift and whether there
is any dependence upon radio luminosity. We find that radio galaxies
appear to belong to a homogeneous population which formed the bulk of
their stars at high redshifts ($z_{\rm f}>5$) and evolved passively
from then until they reach a mean present-day luminosity of
$3\,L_{\star}$. We find a significant difference between the
$K$-magnitudes of the 7CRS and 3CRR radio galaxies with the 7CRS
galaxies being $\approx 0.55$ mag fainter at all redshifts. The cause of
this weak correlation between stellar and radio luminosities probably
lies in mutual correlations of these properties with the central black
hole mass. We compare the evolution-corrected host luminosities at a
constant radio luminosity and find that the typical host luminosity
(mass) increases by approximately $1\,L_{\star}$ from $z \sim 2$ to $z \sim
0.5$ which, although a much smaller factor than predicted by
semi-analytic models of galaxy formation, is in line with results on
optically-selected quasars. Our study has therefore revealed that the
small dispersion in stellar luminosity of radio galaxies around
$3\,L_{\star}$ includes subtle but significant differences between the
host galaxies of extreme- and moderate-power radio sources at fixed
redshift, and between those of high- and low-redshift radio sources at
fixed radio luminosity.

\end{abstract}


\footnotetext{Email:\ chris.willott@nrc.ca}

\section{Introduction}

The luminous radio and narrow-line emission from radio galaxies
provides a convenient method for selecting at least a subset of the
most massive galaxies at a wide range of redshifts. At low redshifts,
powerful radio sources are known to be triggered only within massive
elliptical galaxies (Bettoni et al. 2001). The near-infrared
magnitudes of radio galaxies out to beyond $z=5$ (De Breuck et
al. 2002) suggest that radio galaxies at all redshifts have extreme
stellar masses. Standard CDM-based galaxy formation models predict
that structures grow in a hierarchical fashion. The formation epoch
and evolution of the most massive galaxies is therefore of relevance
to such models.

The near-infrared $K$-band emission from radio galaxies is dominated
by stellar light in most cases (Best, Longair \& R\"ottgering 1998;
Simpson, Rawlings \& Lacy 1999). Although narrow emission lines can
dominate in some objects (Eales \& Rawlings 1993), this is confined to
only the most radio-luminous objects at $z>2$ (e.g. Jarvis et
al. 2001a).  The $K$-band magnitudes of radio galaxies from the bright
3CRR sample (Laing, Riley \& Longair 1983) follow a tight correlation
with redshift (Lilly \& Longair 1984). The nature of this `$K-z$
relation' showed that the stellar luminosities of high redshift radio
galaxies are considerably more luminous than the curve representing no
stellar evolution and consistent with a passively evolving population
which formed at high redshifts. The small dispersion in
$K$-magnitudes at a given redshift evident in the 3CRR sample is also
found in other complete samples and continues up to at least $z=3$
(Jarvis et al. 2001a). Together these facts suggest that radio galaxies
are a homogeneous population which formed the bulk of their stars at
high-redshifts ($z_{\rm f}>5$) and evolved passively from then until
the present day.

Eales et al. (1997) presented $K$-band data for radio galaxies in the
6CE sample which has a flux-density limit a factor of about 5 times lower than
the 3CRR sample. They found that the less radio-luminous 6CE galaxies
tended to have fainter $K$-band magnitudes than 3CRR galaxies at
similar redshifts. They determined a difference of 0.6 mag for the two
samples at $z>0.6$, but found that at $z<0.6$ there was no difference
between the samples. Such an evolutionary change in the
radio-luminosity dependence of stellar luminosities was interpreted by
Best et al. (1998) as due to radio luminosity being more closely
related to the mass of the black hole at high redshifts than at low
redshifts where availability of fuel would dominate. Lacy, Bunker
\& Ridgway (2000) used near-infrared imaging of sources from the
7C--III region of the 7C Redshift Survey (7CRS; selected at a
flux-density limit a factor of 20 times lower than 3CRR) to show that radio
galaxies at $0.8<z<1.4$ from both the 7CRS and 6CE samples are fainter
by 0.4 mag than 3CRR galaxies. They suggest two possible reasons for
this correlation between host and radio luminosities: (i) AGN-related
contamination of the stellar light; (ii) a correlation of both
luminosities with the mass of the central black hole.

It is important in studies of the properties of radio galaxies that
the samples used should be close to complete in terms of secure
optical identifications and redshifts. There are several biases which
can appear for incomplete samples, such as missing objects with the
weakest emission lines because spectra are harder to
obtain. Similarly, spectroscopically targeting the faintest sources at
$K$-band (e.g. with the aim of finding the highest redshift galaxies,
De Breuck et al. 2001) will bias the sample in terms of its
$K-z$ relation.

In this paper we present near-infrared imaging of a
completely-identified sample of radio sources selected at 151 MHz from
the 7C catalogue. This sample (regions 7C--I and 7C--II of the 7CRS)
has 90\% spectroscopic redshift completeness (Willott et al. 2002)
with reasonable estimates of the redshifts for the remaining few
sources from multi-colour photometry (Willott, Rawlings \& Blundell
2001). We combine these data with those from the 3CRR, 6CE (Rawlings
et al. 2001), 6C$^\star$ (Jarvis et al. 2001a,b) and 7C--III (Lacy et
al. 2000) samples to define the $K-z$ relation from over 200 radio
galaxies in samples with high spectroscopic completeness.

In Section 2 we present the $K$-band imaging observations for all the
radio galaxies in 7C--I and 7C--II. In Section 3 we discuss
corrections made to the observed $K$-magnitudes from all the samples
to produce a uniform dataset and then discuss the form of the $K-z$
relation and its evolutionary interpretation. Section 4 investigates
evidence for a correlation between the stellar luminosity (mass) of
the host galaxy and the radio luminosity and also considers the
evolution in stellar luminosity at a fixed radio luminosity.  The
conclusions are given in Section 5. In the Appendix we present
$K$-band images as finding charts for the 7C--I and 7C--II quasars and
broad-lined radio galaxies whose properties were discussed in Willott
et al. (1998). We assume throughout that $H_0=70~ {\rm
km~s^{-1}Mpc^{-1}}$, $\Omega_{\mathrm M}=0.3$ and
$\Omega_\Lambda=0.7$.

\section{Near-infrared imaging}

\subsection{Observations}

All members of the 7C Redshift Survey (except the flat-spectrum quasar
5C\,7.230 and 3C\,200) have been imaged at $K$-band to provide an
identification for spectroscopy, to determine the near-IR morphology
and magnitude. These observations were made over several observing
runs at the United Kingdom Infrared Telescope (UKIRT) on Mauna Kea,
Hawaii. Sky conditions were photometric on all nights. These
observations used the IRCAM3 near-infrared detector. IRCAM3 is a SBRC
InSb $256^{2}$ array with a pixel scale of 0.286 arcsec pix$^{-1}$ and
field of view of $73 \times 73$ arcsec$^{2}$.

In order to subtract the rapidly changing sky background, provide a
good flat-field and avoid problems due to cosmic rays and bad pixels,
the observing strategy used was to offset the position of the
telescope by 10 arcsec between each 60 second exposure. The offsets
were arranged in a $3\times 3$ mosaic. Most sources were observed for
one of these mosaics (total exposure time 9 minutes) or a multiple
thereof. For some of the brightest sources less than 9 minutes was
required to achieve good signal-to-noise. In a few cases, individual
60 second frames were corrupted by the electronics or poor guiding
problems and these were discarded. The observing log for $K$-band
observations of the 7C radio galaxies is given in Table
\ref{tab:obktab}. The observations of quasars and broad-lined radio
galaxies in the sample were described in Willott et al. (1998), but
the images are presented in the Appendix of this paper.

\subsection{Data reduction}

Data reduction was performed using the {\footnotesize IRAF} package.
Dark frames of equal integration time to the science frames were taken
on all nights. An average dark frame was subtracted from each of the
images taken. A flat-field was created for each field observed by
combining all of the data frames from the field using a median filter.
All the dark-subtracted frames were divided by the normalized
flat-field. Registering of the offset frames was performed using the
brightest stars in the images. To correct for the small effects of
atmospheric extinction at $K$-band, each individual frame was scaled
by a factor of $10^{(0.4e\rho)}$, where $e$ is the extinction in
magnitudes at $K$-band per unit airmass and $\rho$ is the airmass
through which the field was observed. A map of the bad pixels in the
array was obtained so that these pixels could be masked out of the
frame combination procedure. All the individual frames of each field
were registered and combined taking the average for each pixel using
the bad pixel mask and a clipping procedure which rejects pixels more
than 4$\sigma$ away from the median of the distribution.  This gives
the final science frame for each field.

\begin{figure*}
\epsfxsize=0.95\textwidth
\begin{centering}
\epsfbox{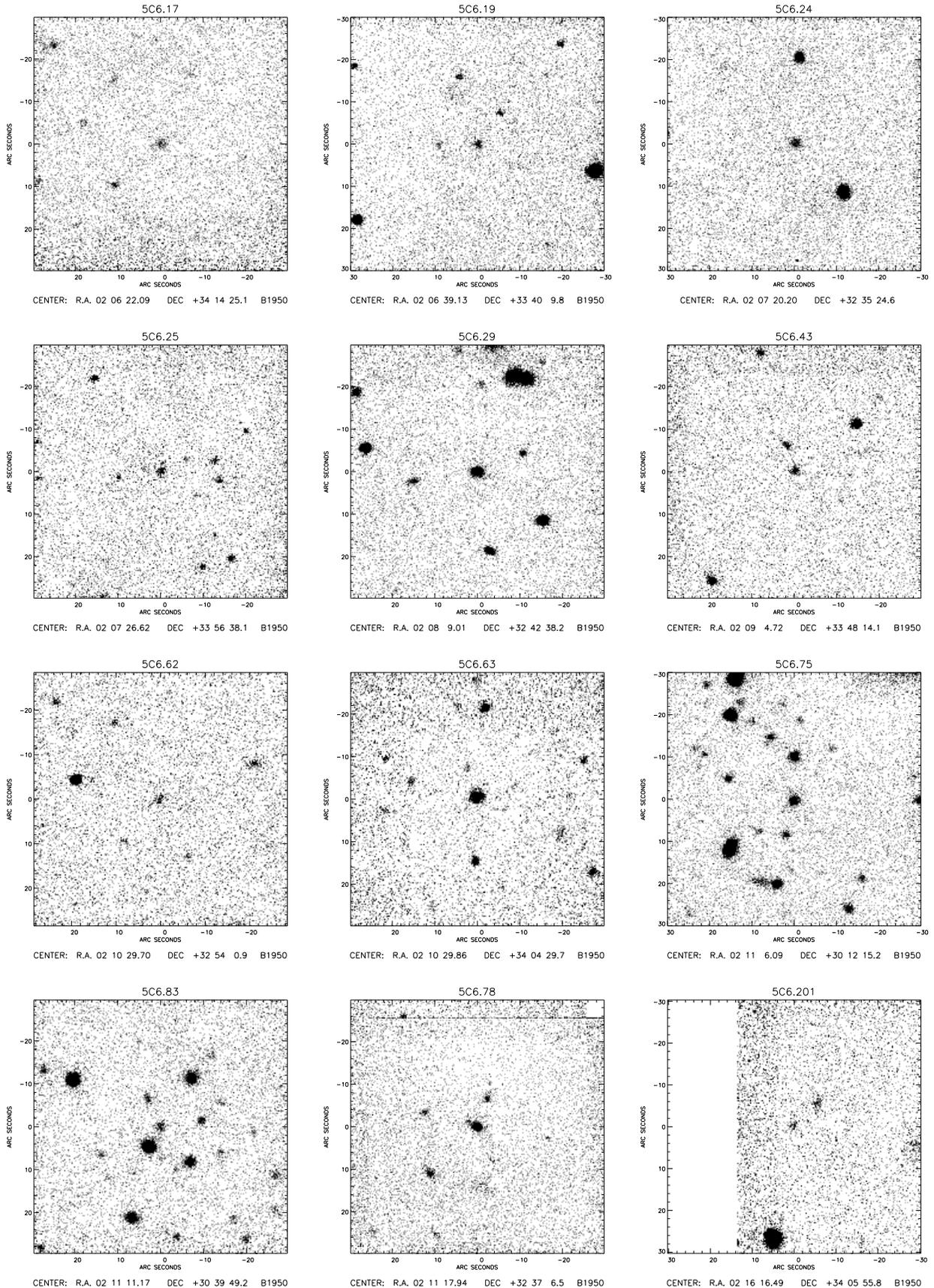} 
\end{centering}
{\caption[junk]{\label{fig:7ckim1}$K$-band images of 7C--I radio galaxies. The images are centred on the near-IR identification of the radio source.}}
\end{figure*}

\addtocounter{figure}{-1}

\begin{figure*}
\epsfxsize=0.95\textwidth
\begin{centering}
\epsfbox{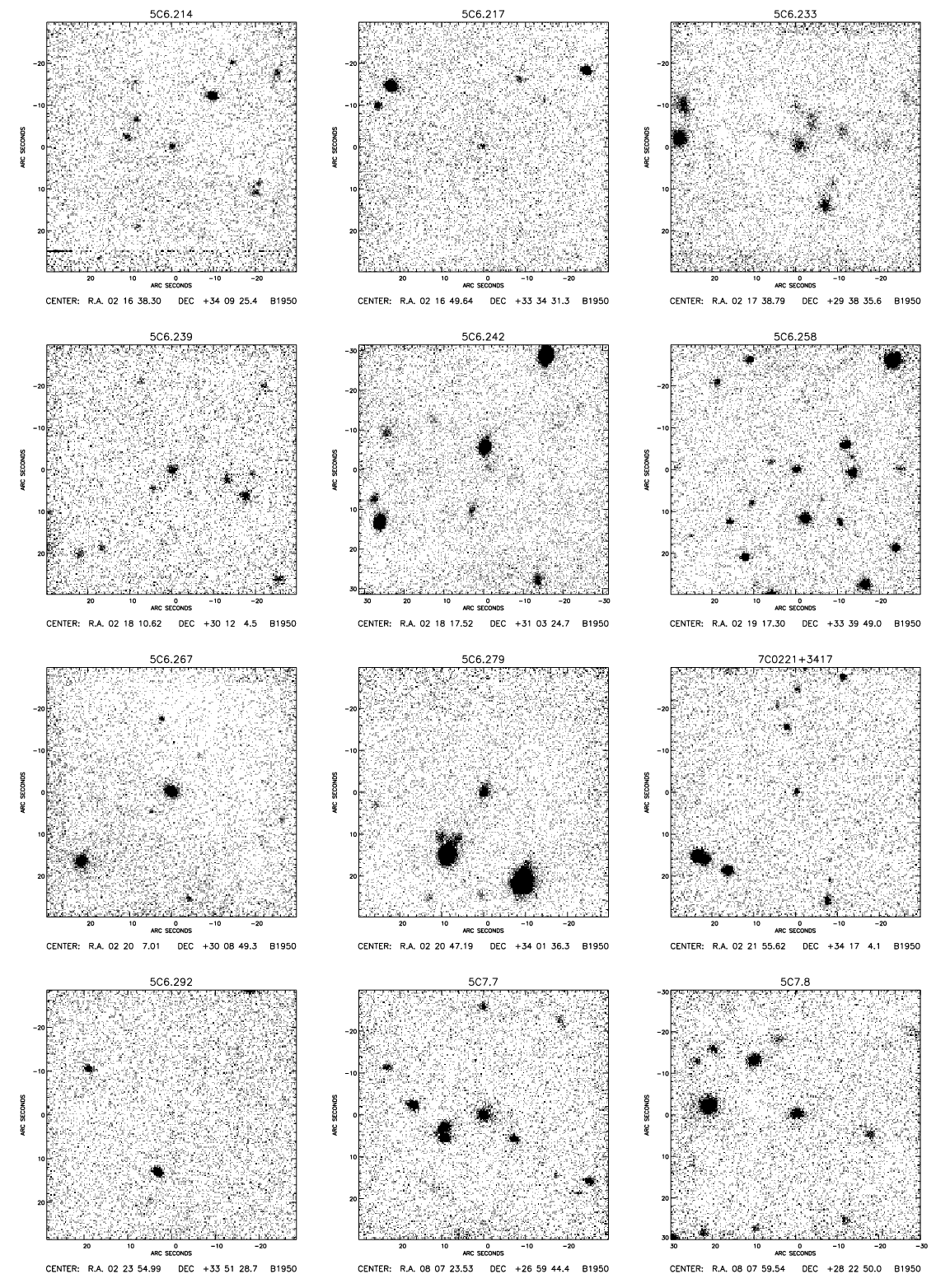} 
\end{centering}
{\caption[junk]{\label{fig:7ckim1} (cont.) $K$-band images of 7C--I/II radio galaxies.}}
\end{figure*}

\addtocounter{figure}{-1}

\begin{figure*}
\epsfxsize=0.95\textwidth
\begin{centering}
\epsfbox{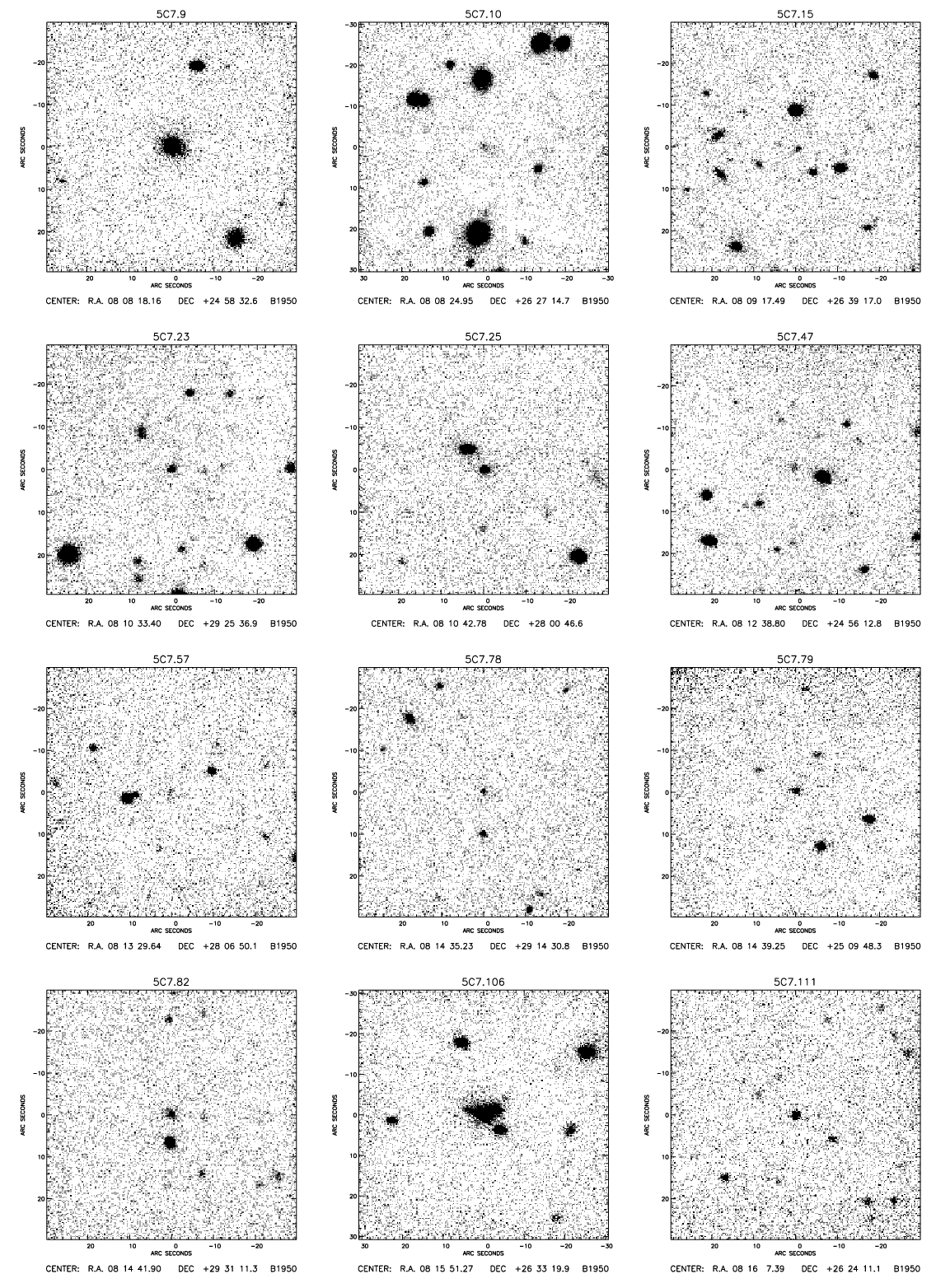} 
\end{centering}
{\caption[junk]{\label{fig:7ckim1} (cont.) $K$-band images of 7C--II radio galaxies.}}
\end{figure*}

\addtocounter{figure}{-1}

\begin{figure*}
\epsfxsize=0.95\textwidth
\begin{centering}
\epsfbox{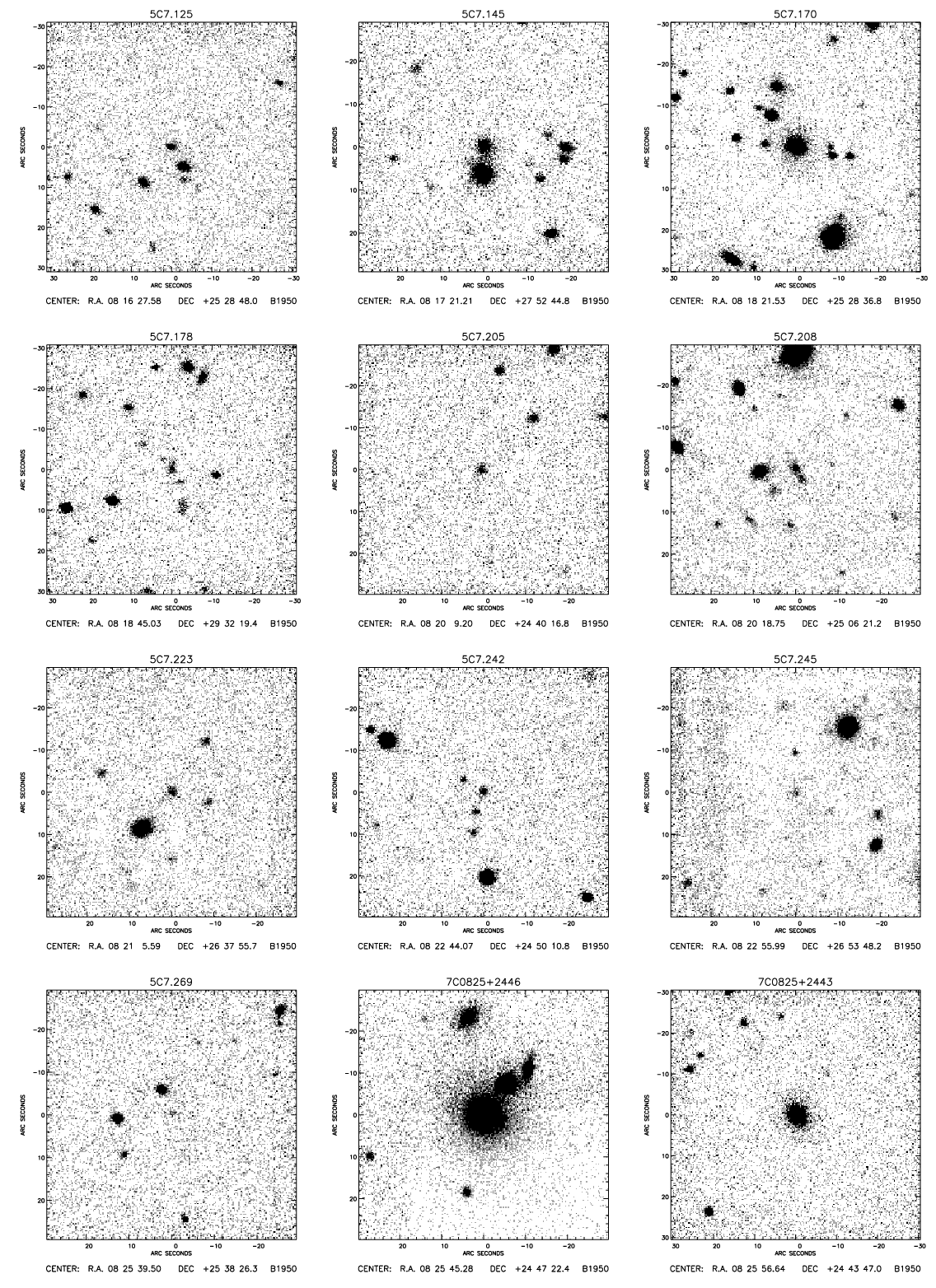} 
\end{centering}
{\caption[junk]{\label{fig:7ckim1} (cont.) $K$-band images of 7C--II radio galaxies.}}
\end{figure*}

\addtocounter{figure}{-1}

\begin{figure}
\hspace{1.0cm}
\epsfxsize=0.95\textwidth
\begin{centering}
\epsfbox{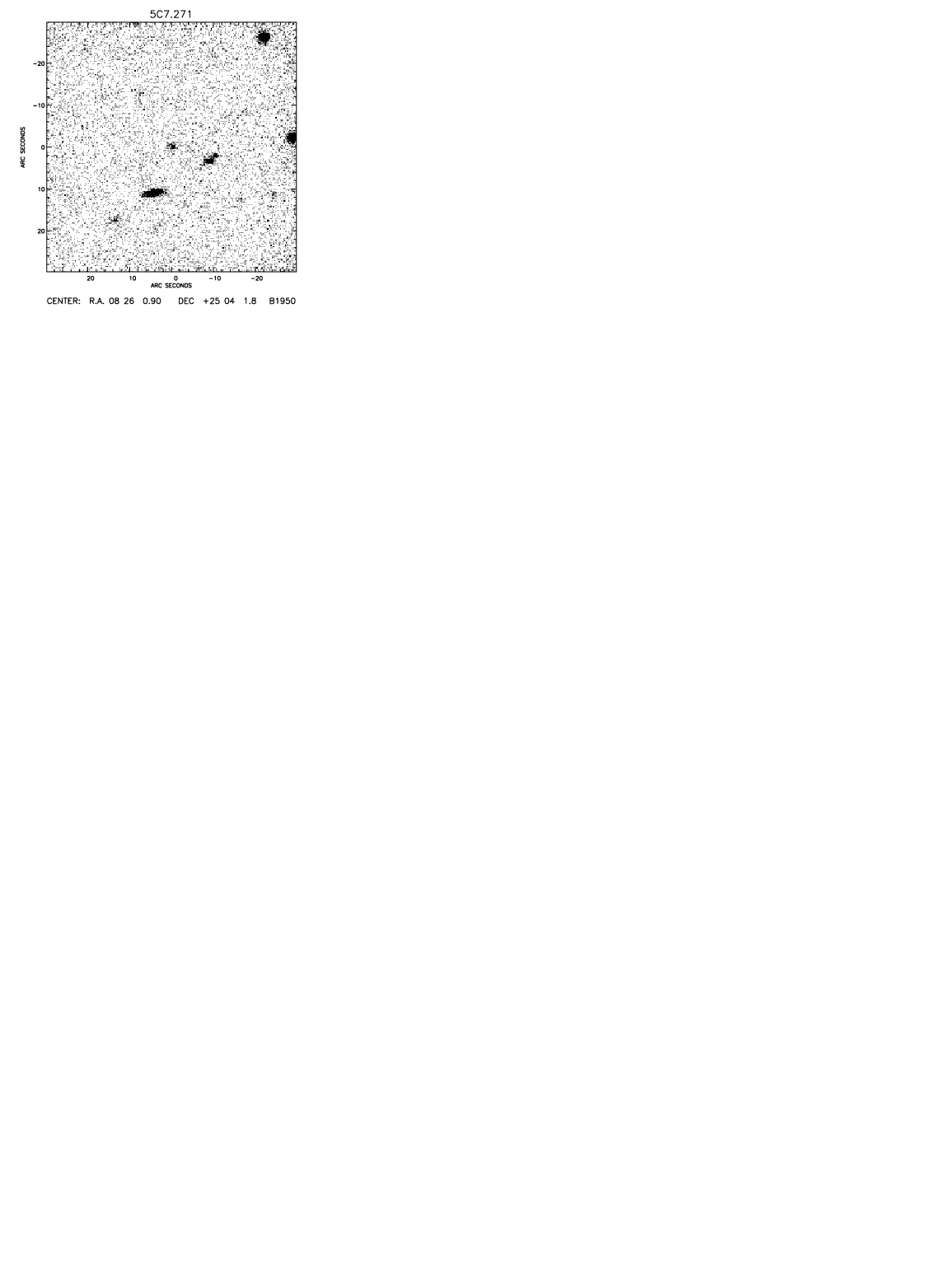} 
\end{centering}
\vspace{-17.7cm}
{\caption[junk]{\label{fig:7ckim1} (cont.) $K$-band image of a 7C--II radio galaxy.}}
\end{figure}

\subsection{Astrometry and identification}

Due to uncertainties on scales of a few arcseconds in the telescope
pointing, it is necessary to accurately determine the astrometry of
the images. This is crucial in this work, since one must ascertain
whether objects detected on the images are the counterparts of the
radio sources. To fix the astrometry of the $K$-band images, finding
charts from the APM catalogue at the Royal Greenwich Observatory,
Cambridge, were obtained for each radio source field (Irwin et
al. 1994). These charts show objects detected on the red POSS plates
to a limit of $R\approx 20$. For most fields, one or more stars
detected on the $K$-band images were also on the APM charts. For cases
where 3 or more stars appear on both the image and the chart, the
{\footnotesize IRAF GASP} package was used to determine the plate
solution for the image. This takes account of any rotation of the
array with respect to North-South (typically of the order of one
degree) and determines the pixel scale. The parameters output by this
task were entered into the headers of the images, so that any pixel
position on the images would be transformed to an accurate position on
the sky. For cases where only one or two APM stars were detected on
the $K$-band image, the plate solution for another image with a good
fit on the same observing run was used along with the position of one
of the detected stars to fix the astrometry. In a few cases no APM
stars were visible on the images, but for these objects wider field
$R$-band images were available and the astrometry could be achieved by
first determining the positions of fainter objects on the $R$-band
images.  

The first step to determining whether the counterparts of the radio
sources were detected on the $K$-band images was by inspection. The
VLA maps resolved the radio sources typically into core and lobe
components (Blundell et al. in prep.). Where a core is clearly
identifiable, it is usual to find the optical counterpart within an
arcsecond or two of this position.  Where no core is visible, the
counterparts are generally found between the two lobes, often
approximately equidistant between the two. Where several objects were
close to the expected counterpart position, the radio and $K$-band
images were overlaid. This enabled a better estimate of the identity
of the true counterpart. Subsequent optical spectroscopy of candidates
revealed the true identification because of the strong emission lines
present in most AGN. Note that in cases where cores and counterparts
were both visible, the typical offset between the two was up to 1
arcsec. This is equal to the estimated internal accuracy of the APM
astrometry (Irwin et al. 1994). Any radio galaxies with ambiguous
identifications are discussed in the accompanying paper on
spectroscopic observations of regions 7C--I and 7C--II of the 7CRS
(Willott et al. 2002).

\begin{table*}
\footnotesize
\begin{center}
\begin{tabular}{llcrcccccc}
\hline\hline
\mc{1}{l}{Name} &\mc{1}{c}{$z$} &\mc{1}{c}{Date}&\mc{1}{c}{Exposure}&\mc{1}{c}{Seeing}&\mc{1}{c}{Optical/NIR Position}&\mc{1}{c}{$K$-mag}&\mc{1}{c}{$K$-mag}&\mc{1}{c}{$K$-mag}\\
\mc{1}{l}{ }      &\mc{1}{c}{}    &\mc{1}{c}{}    &\mc{1}{c}{Time (s)}&\mc{1}{c}{(")}  &\mc{1}{c}{(B1950.0)}                  &\mc{1}{c}{3" aperture}&\mc{1}{c}{5" aperture}&\mc{1}{c}{8" aperture}\\
\hline\hline
5C\,6.17     & 1.05$^{\ddag}$ & 96Feb13  & 1440 & 1.4 &  02 06 22.08~~  +34 14 25.2  & $18.45 \pm 0.08$ &  $17.99 \pm 0.08$ & $17.79 \pm 0.11$ \\
5C\,6.19     & 0.799          & 96Jul29  &  540 & 1.1 &  02 06 39.11~~  +33 40 09.8  & $17.85 \pm 0.08$ &  $17.51 \pm 0.09$ & $17.23 \pm 0.11$ \\ 
5C\,6.24     & 1.073          & 95Jan25  &  540 & 1.5 &  02 07 20.18~~  +32 35 24.8  & $17.74 \pm 0.13$ &  $17.41 \pm 0.12$ & $17.25 \pm 0.13$ \\
5C\,6.25     & 0.706          & 96Jan21  &  540 & 1.1 &  02 07 26.61~~  +33 56 38.3  & $17.88 \pm 0.07$ &  $17.54 \pm 0.07$ & $17.26 \pm 0.09$ \\
5C\,6.29     & 0.720          & 96Jan21  & 1080 & 1.1 &  02 08 09.00~~  +32 42 38.2  & $16.69 \pm 0.02$ &  $16.45 \pm 0.02$ & $16.35 \pm 0.03$ \\
5C\,6.43     & 0.775          & 96Jan21  & 1080 & 1.1 &  02 09 04.70~~  +33 48 14.3  & $17.88 \pm 0.07$ &  $17.69 \pm 0.08$ & $17.76 \pm 0.14$ \\
5C\,6.62     & 1.45$^{\ddag}$ & 97Jan30  &  540 & 1.0 &  02 10 29.69~~  +32 54 01.0  & $18.11 \pm 0.09$ &  $17.70 \pm 0.10$ & $17.41 \pm 0.12$ \\
5C\,6.63     & 0.465          & 97Aug26  &  300 & 1.2 &  02 10 29.85~~  +34 04 29.9  & $16.17 \pm 0.06$ &  $15.84 \pm 0.05$ & $15.71 \pm 0.05$ \\ 
5C\,6.75     & 0.775          & 96Jan21  & 1620 & 1.2 &  02 11 06.09~~  +30 12 15.2  & $17.52 \pm 0.04$ &  $17.22 \pm 0.04$ & $17.06 \pm 0.05$ \\
5C\,6.83     & 1.80$^{\ddag}$ & 97Jan29  & 1080 & 1.3 &  02 11 11.17~~  +30 39 49.3  & $18.30 \pm 0.07$ &  $18.12 \pm 0.10$ &      nbo         \\
5C\,6.78     & 0.263          & 97Jan30  &  180 & 1.0 &  02 11 17.94~~  +32 37 06.6  & $15.95 \pm 0.02$ &  $15.65 \pm 0.03$ & $15.51 \pm 0.04$ \\ 
5C\,6.201    & 0.595          & 97Aug26  &  150 & 1.1 &  02 16 16.49~~  +34 05 56.1  & $17.71 \pm 0.11$ &  $17.47 \pm 0.15$ & $17.45 \pm 0.24$ \\
5C\,6.214    & 0.595          & 97Jan30  &  300 & 1.0 &  02 16 38.28~~  +34 09 25.6  & $17.49 \pm 0.07$ &  $17.20 \pm 0.09$ & $17.25 \pm 0.16$ \\
5C\,6.217    & 1.410          & 96Jul29  &  540 & 1.1 &  02 16 49.62~~  +33 34 31.5  & $18.52 \pm 0.12$ &  $18.34 \pm 0.17$ & $18.02 \pm 0.21$ \\
5C\,6.233    & 0.560          & 95Mar01  & 1080 & 1.4 &  02 17 38.79~~  +29 38 35.6  & $17.83 \pm 0.04$ &  $17.22 \pm 0.04$ & $17.03 \pm 0.06$ \\
5C\,6.239    & 0.805          & 96Jan21  &  540 & 1.2 &  02 18 10.60~~  +30 12 04.5  & $17.48 \pm 0.04$ &  $17.28 \pm 0.05$ & $17.24 \pm 0.08$ \\
5C\,6.242    & 1.90$^{\ddag}$ & 96Feb13  & 2700 & 1.6 &  02 18 17.50~~  +31 03 24.8  & $18.77 \pm 0.07$ &  $18.52 \pm 0.07$ &      nbo         \\ 
5C\,6.258    & 0.752          & 96Jan21  & 1620 & 1.1 &  02 19 17.29~~  +33 39 49.1  & $17.95 \pm 0.04$ &  $17.80 \pm 0.05$ & $17.69 \pm 0.08$ \\
5C\,6.267    & 0.357          & 97Jan30  &  180 & 1.0 &  02 20 07.00~~  +30 08 49.5  & $15.37 \pm 0.01$ &  $15.05 \pm 0.01$ & $14.92 \pm 0.02$ \\
5C\,6.279    & 0.473          & 96Feb13  &  660 & 1.6 &  02 20 47.17~~  +34 01 36.4  & $16.92 \pm 0.03$ &  $16.53 \pm 0.03$ & $16.36 \pm 0.04$ \\
7C\,0221+3417& 0.852          & 97Jan30  &  540 & 1.1 &  02 21 55.60~~  +34 17 04.1  & $18.02 \pm 0.08$ &  $17.76 \pm 0.10$ & $17.48 \pm 0.13$ \\
5C\,6.292    & 1.241          & 96Jul29  &  540 & 1.1 &  02 23 54.99~~  +33 51 28.9  & $19.82 \pm 0.39$ &  $19.27 \pm 0.40$ & $19.37 \pm 0.73$ \\
5C\,7.7      & 0.435          & 96Mar11  &  540 & 1.1 &  08 07 23.52~~  +26 59 44.5  & $16.33 \pm 0.02$ &  $15.96 \pm 0.02$ & $15.73 \pm 0.03$ \\
5C\,7.8      & 0.673          & 95Mar01  &  540 & 1.4 &  08 07 59.52~~  +28 22 50.1  & $16.92 \pm 0.02$ &  $16.61 \pm 0.02$ & $16.48 \pm 0.04$ \\
5C\,7.9      & 0.233          & 96Mar11  &  270 & 1.2 &  08 08 18.14~~  +24 58 32.7  & $14.93 \pm 0.01$ &  $14.54 \pm 0.01$ & $14.34 \pm 0.01$ \\
5C\,7.10     & 2.185          & 95Mar01  & 2700 & 1.2 &  08 08 24.94~~  +26 27 14.8  & $19.44 \pm 0.10$ &  $18.95 \pm 0.10$ & $18.80 \pm 0.15$ \\ 
5C\,7.15     & 2.433          & 96Jan21  &  900 & 1.1 &  08 09 17.49~~  +26 39 17.3  & $18.86 \pm 0.11$ &  $18.56 \pm 0.14$ & $18.69 \pm 0.27$ \\
5C\,7.23     & 1.098          & 95Mar01  & 1620 & 1.1 &  08 10 33.38~~  +29 25 37.2  & $18.34 \pm 0.05$ &  $18.07 \pm 0.06$ & $17.98 \pm 0.09$ \\
5C\,7.25     & 0.671          & 95Jan25  &  540 & 1.2 &  08 10 42.77~~  +28 00 46.6  & $17.16 \pm 0.09$ &  $16.83 \pm 0.08$ & $16.61 \pm 0.08$ \\
5C\,7.47     & 1.70$^{\ddag}$ & 96Jan21  & 4320 & 1.1 &  08 12 38.78~~  +24 56 13.0  & $19.52 \pm 0.10$ &  $19.28 \pm 0.13$ & $19.17 \pm 0.19$ \\  
5C\,7.57     & 1.622          & 96Feb13  &  540 & 1.0 &  08 13 29.64~~  +28 06 50.1  & $19.15 \pm 0.18$ &  $18.59 \pm 0.19$ & $18.76 \pm 0.41$ \\
5C\,7.78     & 1.151          & 97Jan30  &  540 & 0.7 &  08 14 35.21~~  +29 14 31.1  & $18.30 \pm 0.09$ &  $17.91 \pm 0.11$ & $17.75 \pm 0.16$ \\
5C\,7.79     & 0.608          & 96Mar11  &  270 & 1.1 &  08 14 39.24~~  +25 09 48.5  & $17.47 \pm 0.05$ &  $17.18 \pm 0.07$ & $17.12 \pm 0.11$ \\
5C\,7.82     & 0.918          & 97Jan29  &  540 & 1.1 &  08 14 41.90~~  +29 31 11.5  & $17.45 \pm 0.05$ &  $17.06 \pm 0.05$ & $17.01 \pm 0.09$ \\
5C\,7.106    & 0.264          & 96Mar11  &  270 & 1.0 &  08 15 51.26~~  +26 33 20.1  & $15.21 \pm 0.01$ &        nbo        &      nbo         \\
5C\,7.111    & 0.628          & 96Feb13  &  540 & 1.2 &  08 16 07.39~~  +26 24 11.1  & $17.15 \pm 0.03$ &  $17.04 \pm 0.04$ & $16.92 \pm 0.06$ \\ 
5C\,7.125    & 0.801          & 96Mar10  & 1080 & 1.0 &  08 16 27.58~~  +25 28 48.2  & $17.91 \pm 0.04$ &  $17.61 \pm 0.05$ & $17.40 \pm 0.07$ \\ 
5C\,7.145    & 0.343          & 96Mar11  &  270 & 1.1 &  08 17 21.19~~  +27 52 45.0  & $16.06 \pm 0.01$ &  $15.65 \pm 0.01$ & $15.35 \pm 0.02$ \\ 
5C\,7.170    & 0.268          & 96Mar11  &  270 & 1.1 &  08 18 21.53~~  +25 28 36.9  & $15.00 \pm 0.01$ &  $14.57 \pm 0.01$ & $14.33 \pm 0.01$ \\
5C\,7.178    & 0.246          & 96Mar11  &  270 & 1.1 &  08 18 45.02~~  +29 32 19.4  & $17.63 \pm 0.06$ &  $17.36 \pm 0.08$ &      nbo         \\ 
5C\,7.205    & 0.710          & 95Mar01  &  540 & 1.3 &  08 20 09.19~~  +24 40 17.1  & $17.82 \pm 0.05$ &  $17.44 \pm 0.06$ & $17.26 \pm 0.08$ \\
5C\,7.208    & 2.00$^{\ddag}$ & 95Mar01  & 1620 & 1.2 &  08 20 18.74~~  +25 06 21.4  & $18.06 \pm 0.04$ &  $17.60 \pm 0.04$ &      nbo         \\
5C\,7.223    & 2.092          & 95Mar01  & 2700 & 1.2 &  08 21 05.57~~  +26 37 55.8  & $18.76 \pm 0.05$ &  $18.41 \pm 0.06$ & $18.27 \pm 0.09$ \\
5C\,7.242    & 0.992          & 96Feb13  &  480 & 1.1 &  08 22 44.06~~  +24 50 11.0  & $17.60 \pm 0.05$ &  $17.39 \pm 0.06$ & $17.24 \pm 0.09$ \\ 
5C\,7.245    & 1.61$^{\ddag}$ & 95Mar01  & 2700 & 1.2 &  08 22 55.99~~  +26 53 48.2  & $19.19 \pm 0.08$ &  $18.76 \pm 0.09$ & $18.51 \pm 0.11$ \\
5C\,7.269    & 2.218          & 97Jan29  & 1080 & 0.9 &  08 25 39.48~~  +25 38 26.5  & $19.18 \pm 0.14$ &  $18.82 \pm 0.16$ & $18.59 \pm 0.22$ \\
7C\,0825+2446& 0.086          & 97Jan30  &  540 & 1.3 &  08 25 45.26~~  +24 47 22.5  & $13.68 \pm 0.01$ &  $13.09 \pm 0.01$ & $12.72 \pm 0.01$ \\
7C\,0825+2443& 0.243          & 96Mar11  &  270 & 1.1 &  08 25 56.62~~  +24 43 47.0  & $14.97 \pm 0.01$ &  $14.53 \pm 0.01$ & $14.30 \pm 0.01$ \\  
5C\,7.271    & 2.224          & 96Jan21  & 2620 & 1.1 &  08 26 00.89~~  +25 04 01.8  & $18.96 \pm 0.08$ &  $18.77 \pm 0.12$ & $18.73 \pm 0.19$ \\  

\hline
\hline 
\end{tabular} 
\end{center} 

{\caption[Table of observations]{\label{tab:obktab} Log of $K$-band
observations of the 7C--I and 7C--II radio galaxies using IRCAM3 on the
UKIRT. Errors on magnitudes do not include systematic errors due
to zero-point uncertainties (typically $\pm 0.02$ mag). `nbo' in the
aperture magnitude column denotes that for these sources contamination
of the $K$-band flux by a nearby object prevents an accurate
measurement. $^{\ddag}$These redshifts are not from optical
spectroscopy, but have been estimated from optical/near-IR colours and
near-infrared spectroscopy (Willott et al. 2001).

}} 
\normalsize
\end{table*}

\begin{table*}
\begin{center}
\begin{tabular}{lcccclc}
\hline\hline
(1) & (2) & (3) & (4) & (5) & ~(6) & (7)\\
\hline
Name & \mc{1}{c}{$\log_{10} L_{151}$} & \mc{1}{c}{Cl} & \mc{1}{c}{$K$} & 
\mc{1}{c}{$z$} & \mc{1}{l}{Line} & \mc{1}{c}{$\log_{10} L_{\rm line}$} \\
\hline

5C\,6.5      &  26.31  &  Q  &  $16.26 \pm 0.05$ (5)&  1.038 &[OII]\dag&   \f35.62\\
5C\,6.8      &  26.83  &  Q  &  $16.28 \pm 0.06$ (5)&  1.213 &[OIII]&      \f35.85\\
5C\,6.17     &  26.56  &  G  &  $17.79 \pm 0.11$ (8)&  1.05$^{\ddag}$ &H$\alpha$&$<$35.21\\
5C\,6.19     &  26.51  &  G  &  $17.23 \pm 0.11$ (8)&  0.799 &[OII]&       \f34.78\\
5C\,6.24     &  26.68  &  G  &  $17.25 \pm 0.13$ (8)&  1.073 &[OII]&       \f35.55\\
5C\,6.25     &  26.07  &  G  &  $17.26 \pm 0.09$ (8)&  0.706 &[OII]&       \f35.34\\
5C\,6.29     &  25.94  &  G  &  $16.35 \pm 0.03$ (8)&  0.720 &[OII]&       \f35.03\\
5C\,6.33     &  26.64  &  Q  &  $18.23 \pm 0.19$ (5)&  1.496 &[OII]\dag&   \f35.06\\
5C\,6.34     &  27.12  &  Q  &  $15.94 \pm 0.04$ (5)&  2.118 &[OII]\dag&   \f36.70\\
5C\,6.39     &  26.59  &  Q  &  $17.87 \pm 0.08$ (5)&  1.437 &[NeV]&     \f35.37\\
5C\,6.43     &  26.16  &  G  &  $17.76 \pm 0.14$ (8)&  0.775 &[OII]&       \f34.04\\
5C\,6.62     &  26.92  &  G  &  $17.41 \pm 0.12$ (8)&  1.45$^{\ddag}$ &H$\alpha$&$<$35.73\\
5C\,6.63     &  25.66  &  G  &  $15.71 \pm 0.05$ (8)&  0.465 &[OII]&       \f33.95\\
5C\,6.75     &  25.93  &  G  &  $17.06 \pm 0.05$ (8)&  0.775 &[OII]&       \f34.70\\
5C\,6.83     &  27.20  &  G  &  $15.51 \pm 0.04$ (8)&  1.80$^{\ddag}$ &H$\alpha$&$<$35.81\\
5C\,6.78     &  25.62  &  G  &  $18.12 \pm 0.10$ (5)&  0.263 &[OIII]&      \f35.01\\
5C\,6.95     &  27.55  &  Q  &  $16.04 \pm 0.04$ (5)&  2.877 &[OII]\dag&   \f35.91\\
5C\,6.160    &  26.88  &  Q  &  $17.95 \pm 0.14$ (5)&  1.624 &[OIII]&      \f36.58\\
5C\,6.201    &  26.12  &  G  &  $17.45 \pm 0.24$ (8)&  0.595 &[OIII]&     \f 34.39\\
5C\,6.214    &  25.97  &  G  &  $17.25 \pm 0.16$ (8)&  0.595 &[OII]&      \f 34.19\\
5C\,6.217    &  27.13  &  G  &  $18.02 \pm 0.21$ (8)&  1.410 &CIII]&       \f34.91\\
5C\,6.233    &  26.07  &  G  &  $17.03 \pm 0.06$ (8)&  0.560 &[OII]&       \f34.39\\
5C\,6.237    &  27.11  &  Q  &  $15.76 \pm 0.03$ (5)&  1.620 &[OIII]&      \f36.69\\
5C\,6.239    &  26.19  &  G  &  $17.24 \pm 0.08$ (8)&  0.805 &[OII]&       \f35.16\\
5C\,6.242    &  27.06  &  G  &  $18.52 \pm 0.07$ (5)&  1.90$^{\ddag}$ &H$\alpha$&$<$36.09\\
5C\,6.251    &  26.70  &  Q  &  $17.66 \pm 0.11$ (5)&  1.665 &[NeIV]&      \f35.58\\
5C\,6.258    &  25.99  &  G  &  $17.69 \pm 0.08$ (8)&  0.752 &[OII]&       \f34.61\\
5C\,6.264    &  26.27  &  Q  &  $16.16 \pm 0.04$ (5)&  0.831 &[OII]&       \f35.08\\
5C\,6.267    &  25.16  &  G  &  $14.92 \pm 0.02$ (8)&  0.357 &[OII]&       \f34.88\\
5C\,6.279    &  25.54  &  G  &  $16.36 \pm 0.04$ (8)&  0.473 &[OII]&       \f34.83\\
5C\,6.282    &  27.03  &  Q  &  $18.20 \pm 0.18$ (5)&  2.195 &[OII]\dag&   \f35.52\\
7C\,0221+3417&  26.79  &  G  &  $17.48 \pm 0.13$ (8)&  0.852 &[OII]&       \f34.69\\
5C\,6.286    &  26.65  &  Q  &  $17.62 \pm 0.20$ (3)&  1.339 &[OIII]&   $<$35.35\\
5C\,6.288    &  27.60  &  Q  &  $18.28 \pm 0.12$ (4)&  2.982 &Ly$\alpha$& \f 36.68\\
5C\,6.287    &  27.57  &  Q  &  $16.11 \pm 0.04$ (5)&  2.296 &[OII]\dag&  \f 36.78\\
5C\,6.291    &  27.56  &  Q  &  $16.06 \pm 0.03$ (5)&  2.910 &Ly$\alpha$& \f 37.10\\
5C\,6.292    &  26.72  &  G  &  $19.37 \pm 0.73$ (8)&  1.241 &[OII]&      \f 35.93\\

 \hline\hline
 \end{tabular}
 {\caption[Table 1]{\label{tab:summary5c6} A summary of key
information on the 7C--I sample.  Full details of spectroscopic
observations are given in Willott et al. (1998) for the quasars and
broad-lined radio galaxies and Willott et al. (2002) for the
narrow-line radio galaxies. Note that the cosmological parameters
assumed here to calculate the radio and emission line luminosities
differ from those used in many of our previous papers.  {\bf Column
1:} Name of the radio source. Note that most of the names begin with
5C\,6 because they were originally catalogued in the 5C survey in this
region (Pearson \& Kus 1978).
{\bf Column 2:} $\log_{10}$ of the rest-frame 151-MHz radio luminosity
in units of W\,Hz$^{-1}$\,sr$^{-1}$.
{\bf Column 3:} Classification, Q=quasar, B=broad-line radio galaxy,
G=narrow-line radio galaxy following the prescription of Willott et
al. (1998).
{\bf Column 4:} $K$-band magnitude of the radio source ID and
photometric error. The number in brackets is the diameter in arcsec of
the photometric aperture.
{\bf Column 5:} Redshift. $^{\ddag}$These redshifts are not from optical
spectroscopy, but have been estimated from optical/near-IR colours and
near-infrared spectroscopy (Willott et al. 2001)
{\bf Column 6:} Prominent narrow emission line in the existing
spectra. \dag indicates that the line has not been observed for this
quasar but the line luminosity has been estimated assuming a typical
[OII] equivalent width of 10 \AA.
{\bf Column 7:} $\log_{10}$ of the line luminosity in units of W, a
$<$ denotes that the line is not detected and an upper limit is
quoted. 

}}

 \end{center}
 \end{table*}




\begin{table*}
\begin{center}
\begin{tabular}{lcccclc}
\hline\hline
(1) & (2) & (3) & (4) & (5) & ~(6) & (7)\\
\hline
Name & \mc{1}{c}{$\log_{10} L_{151}$} & \mc{1}{c}{Cl} & \mc{1}{c}{$K$} & 
\mc{1}{c}{$z$} & \mc{1}{l}{Line} & \mc{1}{c}{$\log_{10} L_{\rm line}$} \\
\hline

5C\,7.7      &  25.57  &  G  &  $15.73 \pm 0.03$ (8)&  0.435 &[OII]&       \f34.28\\
5C\,7.8      &  26.36  &  G  &  $16.48 \pm 0.04$ (8)&  0.673 &[OII]&       \f34.97\\
5C\,7.9      &  25.36  &  G  &  $14.34 \pm 0.01$ (8)&  0.233 &[OIII]&      \f35.52\\
5C\,7.10     &  27.54  &  G  &  $18.80 \pm 0.15$ (8)&  2.185 &Ly$\alpha$&  \f36.95\\
7C\,0808+2854&  27.13  &  Q  &  $16.06 \pm 0.04$ (5)&  1.883 &[OIII]&      \f36.76\\
5C\,7.15     &  27.35  &  G  &  $18.69 \pm 0.27$ (8)&  2.433 &Ly$\alpha$&  \f36.25\\
5C\,7.17     &  26.20  &  B  &  $17.37 \pm 0.07$ (5)&  0.936 &[OII]&       \f35.25\\
5C\,7.23     &  26.63  &  G  &  $17.98 \pm 0.09$ (8)&  1.098 &[OII]&       \f35.16\\
5C\,7.25     &  25.77  &  G  &  $16.61 \pm 0.08$ (8)&  0.671 &[OII]&       \f34.77\\
5C\,7.47     &  26.79  &  G  &  $19.17 \pm 0.19$ (8)&  1.70$^{\ddag}$ &MgII&     $<$34.98\\
5C\,7.57     &  26.78  &  G  &  $18.76 \pm 0.41$ (8)&  1.622 &CIII]&       \f35.02\\
5C\,7.70     &  27.75  &  Q  &  $17.48 \pm 0.09$ (5)&  2.617 &[OII]\dag&   \f36.02\\
5C\,7.78     &  26.99  &  G  &  $17.75 \pm 0.16$ (8)&  1.151 &[OII]&       \f35.47\\
5C\,7.79     &  25.76  &  G  &  $17.12 \pm 0.11$ (8)&  0.608 &[OII]&       \f34.24\\
5C\,7.82     &  26.28  &  G  &  $17.01 \pm 0.09$ (8)&  0.918 &[OII]&       \f34.86\\
5C\,7.85     &  26.63  &  Q  &  $16.24 \pm 0.03$ (5)&  0.995 &[OII]&       \f35.18\\
5C\,7.87     &  27.17  &  Q  &  $18.74 \pm 0.22$ (5)&  1.764 &[OII]\dag&   \f35.21\\
5C\,7.95     &  26.65  &  Q  &  $16.61 \pm 0.03$ (5)&  1.203 &[OIII]&      \f36.17\\
5C\,7.106    &  25.27  &  G  &  $15.21 \pm 0.01$ (3)&  0.264 &[OIII]&   $<$33.82\\
5C\,7.111    &  26.29  &  G  &  $16.92 \pm 0.06$ (8)&  0.628 &[OII]&       \f34.70\\
5C\,7.118    &  26.05  &  B  &  $15.75 \pm 0.03$ (5)&  0.527 &[OIII]&      \f34.97\\
5C\,7.125    &  26.14  &  G  &  $17.40 \pm 0.07$ (8)&  0.801 &[OII]&       \f34.63\\
5C\,7.145    &  25.31  &  G  &  $15.35 \pm 0.02$ (8)&  0.343 &[OII]&    $<$33.94\\
5C\,7.170    &  25.18  &  G  &  $14.33 \pm 0.01$ (8)&  0.268 &[OIII]&   $<$33.68\\
5C\,7.178    &  25.14  &  G  &  $17.36 \pm 0.08$ (5)&  0.246 &[OIII]&      \f32.85\\
5C\,7.194    &  27.29  &  Q  &  $15.91 \pm 0.03$ (5)&  1.738 &[NeIV]&      \f36.07\\
5C\,7.195    &  27.12  &  Q  &  $17.71 \pm 0.05$ (5)&  2.034 &Ly$\alpha$&  \f37.04\\
5C\,7.205    &  26.34  &  G  &  $17.26 \pm 0.08$ (8)&  0.710 &[OII]&       \f35.20\\
5C\,7.208    &  27.27  &  G  &  $17.60 \pm 0.04$ (5)&  2.00$^{\ddag}$ &H$\alpha$&$<$35.61\\
5C\,7.223    &  27.06  &  G  &  $18.27 \pm 0.09$ (8)&  2.092 &Ly$\alpha$&  \f36.71\\
5C\,7.242    &  26.21  &  G  &  $17.24 \pm 0.09$ (8)&  0.992 &[OII]&       \f34.71\\
5C\,7.245    &  27.22  &  G  &  $18.51 \pm 0.11$ (8)&  1.61$^{\ddag}$ &H$\alpha$&   \f35.92\\
7C\,0825+2930&  26.95  &  Q  &  $17.55 \pm 0.15$ (4)&  2.315 &[OII]\dag&   \f35.96\\
5C\,7.269    &  27.21  &  G  &  $18.59 \pm 0.22$ (8)&  2.218 &Ly$\alpha$&  \f36.51\\
7C\,0825+2446&  24.85  &  G  &  $12.72 \pm 0.01$ (8)&  0.086 &[OIII]&   $<$32.82\\
7C\,0825+2443&  24.93  &  G  &  $14.30 \pm 0.01$ (8)&  0.243 &[OIII]&   $<$33.45\\
5C\,7.271    &  27.06  &  G  &  $18.73 \pm 0.19$ (8)&  2.224 &CIII]&       \f34.96\\

\hline\hline
\end{tabular}
{\caption[Table 1]{\label{tab:summary5c7} A summary of key information
on the 7C--II sample in the same format as Table
\ref{tab:summary5c6}. Most of the names begin with 5C\,7 because they
were originally catalogued in the 5C survey in this region (Pearson \&
Kus 1978). Note that the 3CRR source 3C~200 should technically be
included in the 7C-II sample, since it is contained within this region
of sky and has a 151 MHz flux-density above the limit. }}

\end{center}
\end{table*}

\subsection{Photometry}

Aperture photometry of radio source identifications was performed
using the {\footnotesize IRAF APPHOT} package. The reduction procedure
described above does not subtract the background level off the final
images. However, the flat-fielding ensures that there is little
spatial structure in the background level. To determine the $K$
magnitudes of detected objects, the local background near each object
was measured by averaging over the pixels in an annulus about the
object. The inner radius and width of the sky annulus were 5 and 8
arcsec, respectively. Note that any pixels above 3$\sigma$ of the
median counts in the sky region are rejected, ensuring that nearby
sources do not contaminate the background level. Photometry was
performed using three different diameter apertures of 3, 5 and 8
arcsec.

Over the course of each night, several UKIRT faint standard stars were
observed (Casali \& Hawarden 1992) for photometric calibration. The
reduction process used for these standard star observations was
identical to that outlined above for the object frames. Aperture
photometry of the standard stars gave the zero-point for the
images. The zero-points obtained remained fairly constant throughout
individual nights, typically with differences of only $\pm 0.02$ mag. The
zero-points obtained at $K$-band were $\approx 22.5$ mag for all the
observing runs. The $K$-magnitudes and positions of radio source
identifications are given in Table \ref{tab:obktab}. Fig.\
\ref{fig:7ckim1} shows greyscale representations of the $K$-band
images. A summary of the basic information on all the sources in
regions 7C--I and 7C--II of the 7CRS is given in Tables
\ref{tab:summary5c6} and \ref{tab:summary5c7}.

\section{The $K-z$ relation}

\subsection{Data from other samples, and aperture and emission lines corrections}

The acquisition of complete $K$-band photometry of this sample of
radio sources together with $>90\%$ spectroscopic redshifts enables us
to examine the evolution of the stellar populations of radio
galaxies. We do not include in this paper objects which show broad
emission lines since they can have a high level of contamination of
the $K$-band light by non-stellar quasar emission. We do not impose
any criteria on the radio structures of the sources considered here,
so both FRI and FRII type sources are included. Radio galaxies at
$z<0.05$ are excluded from our analysis because their large apparent
sizes lead to very large aperture corrections from aperture
magnitudes.

Lacy et al. (2000) present near-IR data for all radio galaxies at
$z>0.8$ in the 7C--III region of the 7CRS. Data on all 7C--III
galaxies at $z>1.2$ are also included here with the 7C--I and 7C--II
data described in Sec.\ 2. This redshift limit was chosen to avoid
including objects with data only in the $J$-band for which large
colour corrections would be necessary to convert to $K$-band. Only two
of the sixteen sources at $z>1.2$ do not have $K$-band data and for
these objects at $z \approx 1.5$ we assume a colour of $H-K=0.9$ to
convert from the $H$-band magnitudes in Lacy et al.

In addition we can combine our data with those from brighter complete
samples to search for any correlation of near-IR luminosity with radio
luminosity. The two complete flux-density limited samples selected at
a similar radio frequency to the 7CRS are the 3CRR and 6CE
samples. $K$-band photometry of 3CRR sources is taken from Lilly \&
Longair (1984), Best et al. (1998), Simpson, Ward \& Wall (2000), de
Vries et al. (1998), Hutchings \& Neff (1997), Rawlings et al. (1996)
and Stockton, Canalizo \& Ridgway (1999) using the more recent data
where multiple measurements exist. Note that this does not include
every radio galaxy in the 3CRR sample. 27 out of the 96 3CRR radio
galaxies with $z \ge 0.05$ do not have $K$-band photometry in these
publications. We assume that the observed sample is representative and
that no biases are introduced by the missing sources.  A compilation
of data for the complete 6CE sample is given in Rawlings et
al. (2001). We also include data from the filtered 6C$^\star$ sample
because the linear size and spectral index criteria do not seem to
affect the position on the $K-z$ relation and this sample has very
high spectroscopic redshift completeness (Jarvis et al. 2001a,b). The
total numbers of $z \ge 0.05$ radio galaxies in each of the samples
with $K$-band data are 69 in 3CRR, 47 in 6CE, 65 in 7CRS and 23 in
6C$^\star$ (total 204). The data for all the samples used in this
paper are available on-line at {\bf
http://www-astro.physics.ox.ac.uk/$\sim$cjw/kz/kz.html}.

\begin{figure*}
\epsfxsize=0.95\textwidth
\begin{centering}
\epsfbox{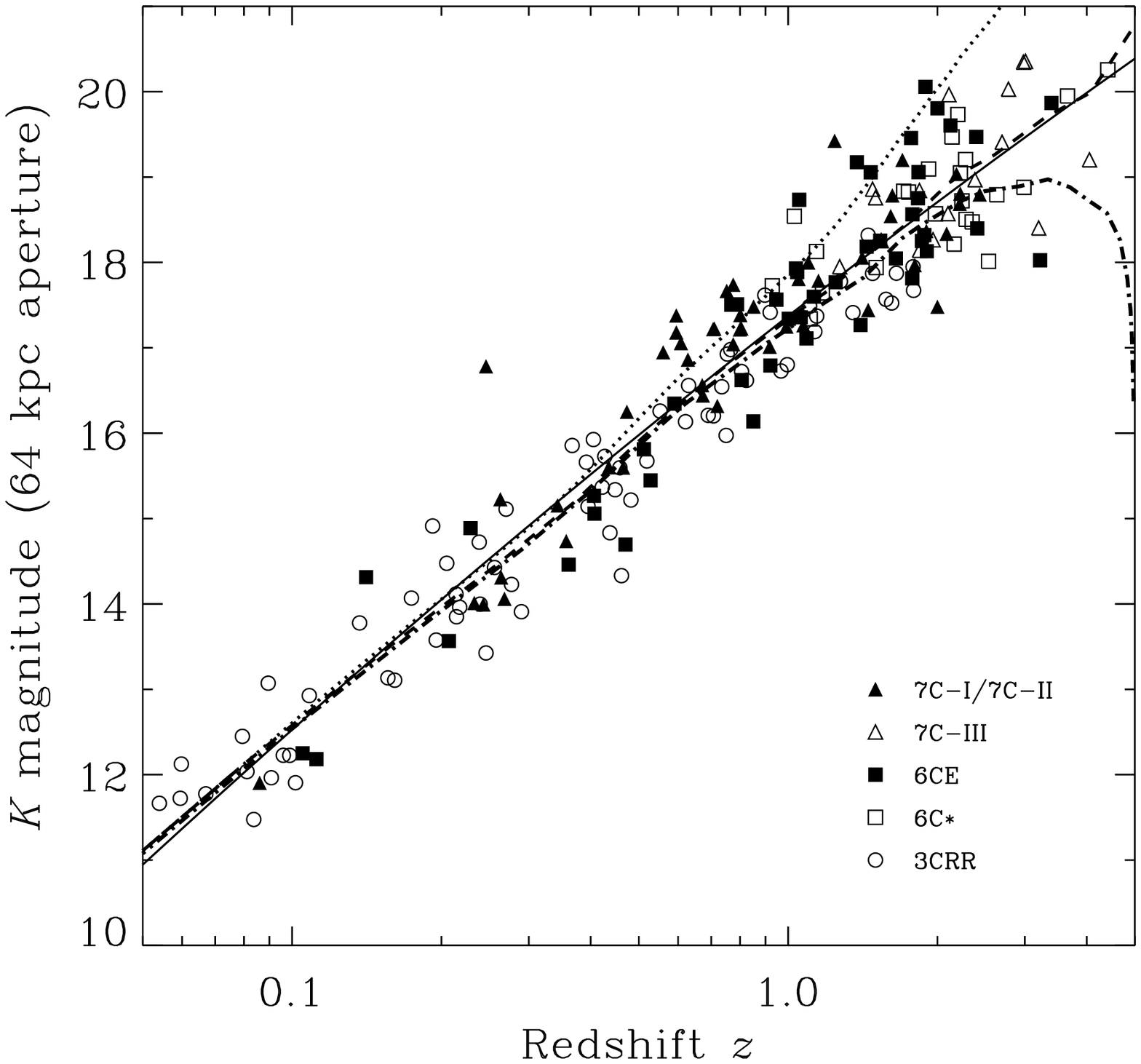} 
\end{centering}

{\caption[junk]{\label{fig:kzall} Aperture and emission line corrected
$K$-band magnitude versus redshift for radio galaxies from the 7CRS,
6CE, 6C$^\star$ and 3CRR complete samples. The solid curve is the
best-fit second order polynomial relationship between $K$ and
$\log_{10}z$ for all the data. This is remarkably similar to the
apparent $K$-magnitude evolution of a galaxy of local luminosity
$3\,L_{\star}$ which forms all its stars in an instantaneous burst at
$z_{\rm f}=10$ (dashed line). A similar model with $z_{\rm f}=5$
(dot-dashed line) predicts galaxies are much more luminous at $z>3$. The
no-evolution model for a $3\,L_{\star}$ galaxy (dotted) provides a poor
fit to the $z>1$ data.

}}

\end{figure*}

$K$-band magnitudes of radio galaxies in the 7C--I, 7C--II and 6C$^\star$
samples are quoted in 3,5 and 8 arcsec apertures. However, for the
3CRR, 6CE and 7C--III radio galaxies the data are given in various
apertures. We apply the same procedure as in Jarvis et al. (2001a) to
convert from apparent angular size aperture magnitudes to standard
64 kpc metric apertures to be consistent with previous studies. As
highlighted by Eales et al. (1993), the strong emission lines of the
most powerful radio sources can provide significant flux contamination
at $K$-band, particularly at $z\approx 2$ where the H$\alpha$ line
lies in the $K$-band. Jarvis et al. (2001a) accounted for the emission
line contribution by assuming that sources follow the emission line --
radio correlation of Willott (2001). We adopt this method in this
paper and correct the $K$-magnitudes of all sources for the expected
emission line flux.

\subsection{The evolution of the stellar hosts of radio galaxies}
\label{kz}

In Fig.\ \ref{fig:kzall} we show the near-infrared Hubble diagram of
$K$-magnitude versus $z$ for the 7CRS, 6CE, 6C$^\star$ and 3CRR
samples. The solid line is the best-fit second order polynomial
relationship between $K$ and $\log_{10}z$ for all the data (calculated
by minimizing the square of the residuals in $K$). Although in Sec.\
\ref{lumdep} we will show that the different samples do follow
slightly different $K-z$ relations, the weakness of this effect and
the relative lack of evolution of the difference means that we can
include all the data here to define the $K-z$ relation for all
powerful radio galaxies. The best-fit relation is
\begin{equation}
K=17.37+4.53 \log_{10} z - 0.31 (\log_{10} z)^{2}
\end{equation}

Eales et al. (1997) found that the dispersion in the $K-z$ relation
for the 6CE sample increased markedly at redshifts $z>2$ and proposed
that this was revealing the epoch of formation of massive
ellipticals. A similar effect was found by Lacy et al. (2000)
including the 7C--III data and high redshift radio galaxies from the
literature.  However Jarvis et al. (2001a) found no increase in the
dispersion up to $z \sim 3$ for the combination of the 6CE and
6C$^\star$ samples.  We have evaluated the dispersion about our
best-fit relation as a function of redshift. We find a standard
deviation $\sigma$ that is approximately constant at all redshifts:
$\sigma=0.57$ at $0.05<z<1$, $\sigma=0.60$ at $1<z<2$, $\sigma=0.59$
at $2<z<3$. This appears to be in agreement with the results of Jarvis
et al. (2001a), but we caution that the highest redshift bin does not
include any 3C sources so the correlation between radio luminosity and
$K$-magnitude (Sec.\ \ref{lumdep}) may be increasing the dispersion in
the lower redshift bins (although Jarvis et al. find comparable
dispersions in the 6C data alone at these lower redshifts). One of the
7CRS galaxies at $z=0.246$ is much fainter than all other radio
galaxies at this redshift and is the only major outlier from the $K-z$
relation. This source, 5C\,7.178, is discussed further in Willott et
al. (2002).

Also plotted on Fig.\ \ref{fig:kzall} are curves showing the evolution
in apparent $K$-magnitude for several model galaxies determined using
GISSEL96 (Bruzual \& Charlot 2002). The upper curve (dotted line) is
for a model galaxy with luminosity $3\,L_\star$ (assuming $M^\star_K=-24.3$
for local ellipticals from Kochanek et al. 2000) which includes the
$k$-correction but no evolution in the stellar population. At all
redshifts, the spectrum of the model galaxy is that of a 13 Gyr old
instantaneous burst with solar metallicity (note that this violates
constraints from the age of the Universe at high redshifts). As first
shown by Lilly \& Longair (1984), at $z>1$, radio galaxies are
typically more luminous than their local counterparts, implying significant
evolution in their stellar populations. This is evident in Fig.\
\ref{fig:kzall} by the poor fit of the non-evolving model.

We have also evaluated the evolution in $K$-magnitude for evolving
stellar population models which formed all their stars in an
instantaneous burst at high redshift. Evolving models for a
$3\,L_\star$ galaxy with formation redshifts of $z_{\rm f}=5$
(dot-dashed) and $z_{\rm f}=10$ (dashed) are also plotted on Fig.\
\ref{fig:kzall}.  The $z_{\rm f}=10$ curve is remarkably close to the
best-fit relation over the whole redshift range, indicating the mean
luminosity of radio galaxies is $3\,L_\star$\footnotemark. However at
least some of the $z>3$ radio galaxies lie closer to the $z_{\rm f}=5$
curve. It is also worth recalling that in modern theoretical models
for galaxy formation (e.g. Somerville \& Primack 1999; Cole et
al. 2000), there is a significant amount of merging at
high-redshifts. This means that objects which appear to follow the
$z_{\rm f}=10$ passive evolution curve could equally well be less
massive galaxies with a more recent star-formation epoch. We will not
discuss the evolutionary implications of the $K-z$ relation in detail
here, since these issues have recently been discussed for subsets of
these data by Jarvis et al. (2001a) and Inskip et al. (2002). We note
that Inskip et al. stated that the passive evolution model with
$z_{\rm f}=10$ in a $\Omega_{\rm M}=0.3,\Omega_\Lambda=0.7$ cosmology
did not provide a good fit to their combined 3CRR and 6CE sample. The
deviations of their model from the binned data were only $\approx 0.4$
magnitudes at $z=0.5, z=1$ and $z=2$, i.e. less than the standard
deviation in the data and formally consistent with the evolution over
the redshift range $0.5<z<2$. Their differing result from ours is due
to the new data presented here, our emission line corrections and a
slightly different low-redshift normalization. We conclude that the
passive evolution model is consistent with the $K-z$ relation in a
$\Lambda$-dominated cosmology, especially given the further
complications to be discussed in Sec.\ \ref{lumdep}.

\footnotetext{Jarvis et al. (2001a) used a similar dataset and the
same cosmological model as in this paper and quoted the mean $K$-band
luminosity of radio galaxies as $5\,L_\star$. This was due to their
use of $M^\star_K$ from Gardner et al. (1997), although the text
states that the value of $M^\star_K$ used was from the more recent
derivation by Kochanek et al. (2000). This does not affect any of
their other results.}

\section{Radio-luminosity dependence of the stellar host luminosity}
\label{lumdep}

There is a simple explanation of why we may expect to find that the
stellar luminosities of radio galaxies are correlated with radio
luminosity. In essence, this comes down to the fact that more massive
objects are generally more luminous. In this case there are several
observed correlations that can be used to specify the relationship
between stellar and radio luminosity. First, the stellar luminosity
of ellipticals is known to correlate almost linearly with the central
black hole mass (Magorrian et al. 1988), presumably because the
stellar luminosity is very closely related to the stellar mass and
there is a tight correlation between host and black hole masses
(Gebhardt et al. 2000; Ferrarese \& Merritt 2000). The radio
luminosity is found to be linearly related to the narrow emission line
luminosity (Willott 2001) and therefore presumably to the UV ionizing
luminosity (Rawlings \& Saunders 1991; Willott et al. 1999). The UV
ionizing luminosity draws its power from the accretion of material
onto the supermassive black hole. Therefore any correlation between
stellar and radio luminosities is likely to have its cause in the fact
that both properties correlate positively with the black hole
mass. This means that by measuring the strength of any correlation
between the stellar and radio luminosities, we can estimate the range
of black hole masses in high-redshift radio galaxies.

\begin{figure}
\epsfxsize=0.48\textwidth
\begin{centering}
\epsfbox{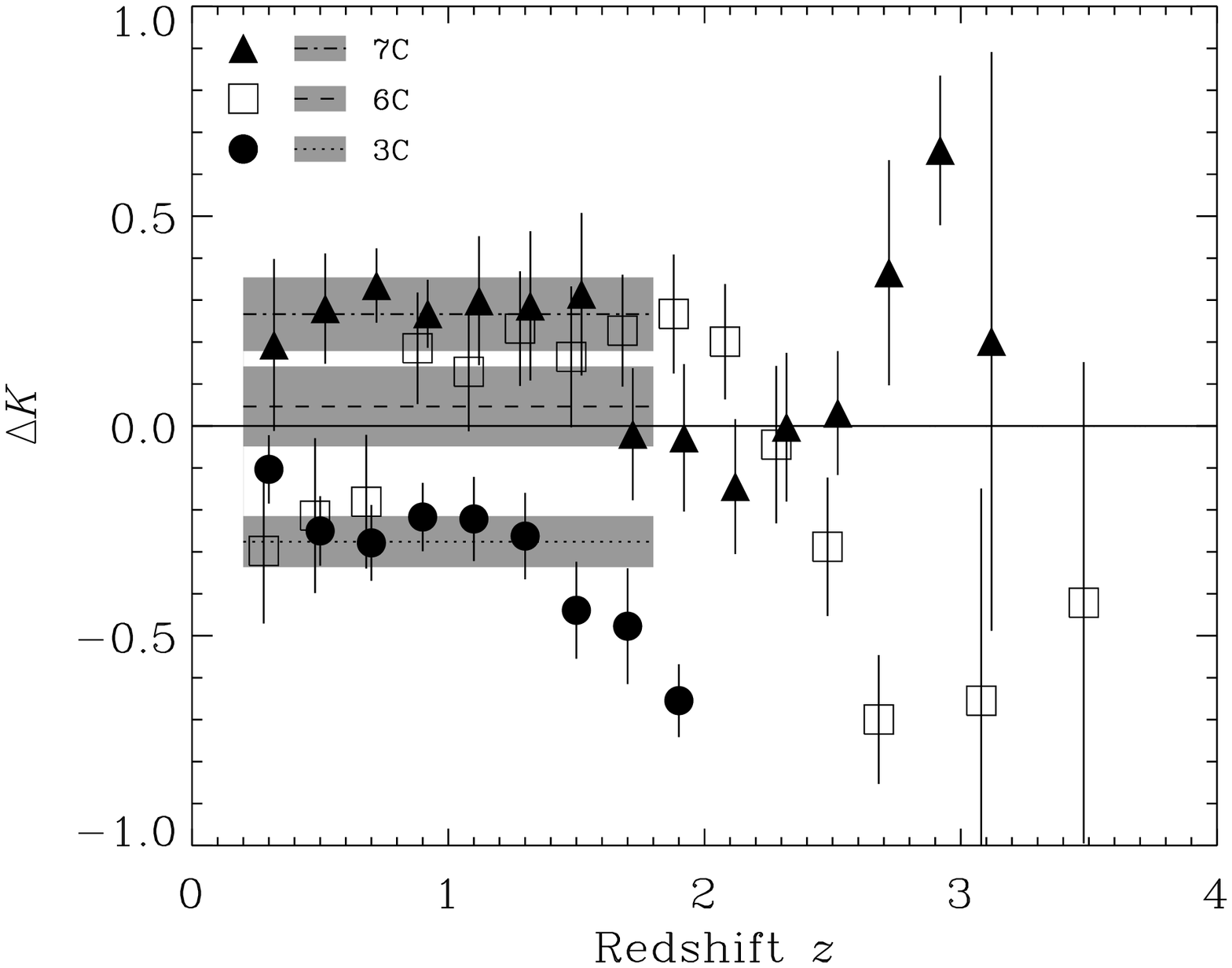} 
\end{centering}

{\caption[junk]{\label{fig:kzres} Residuals from the best fit to the
$K-z$ relationship ($\Delta K= K-K_{\rm fit}$) against redshift. The
data for each sample have been binned in bins of width $\Delta z =0.6$
and the error bars show the standard errors which are correlated as
described in Sec.\ \ref{lumdep}. Only bins containing 3 or more
sources are plotted. The 6C and 7C bins are shown offset from their
true central redshifts by $\pm 0.02$ for clarity.  The dot-dashed,
dashed and dotted lines show the mean value of $\Delta K$ over the
redshift range $0.2<z<1.8$ for the 7C, 6C and 3C samples,
respectively. The grey shading about these lines show the $\pm 1
\sigma$ standard errors about the means.}}

\end{figure}

To investigate the correlation between the stellar and radio
luminosities we determine the residuals for each object about the best
fitting $K-z$ relation presented in Sec.\ \ref{kz} and Fig.\
\ref{fig:kzall}. We do not bin the sources in terms of radio
luminosity since the sources from different samples span a range of
redshifts at a given luminosity and we wish to investigate the
evolutionary dependence of any correlation.  However, because the
samples are radio flux-limited the median flux-density of 6C sources is
$\approx 7$ times fainter than that of 3C sources and the 7C sources a
further factor of $\approx 2.5$ times fainter than 6C. Therefore the
radio luminosities of sources at a given redshift typically differ by
these amounts and we determine the mean residual for each sample as a
function of redshift.  To obtain a large enough number of objects per
sample per bin, the bins in redshift have width $\Delta z=0.6$. To
avoid spurious results due to the choice of binning, we have evaluated
the mean residual per sample per bin for a range of bins from
$0<z<0.6$ to $3.2<z<3.8$ with steps of 0.2 in $z$. This does mean that
neighbouring points are correlated but gives a more accurate measure
of the true differences between the samples as a function of epoch.

In Fig.\ \ref{fig:kzres} we plot the mean $K$-magnitude residuals as a
function of redshift for the three sets of samples: 7C (7C--I, 7C--II and
7C--III), 6C (6CE and 6C$^\star$) and 3C (3CRR). The first thing to
notice is a statistically significant difference (at the $>95$\%
level) between the 7C and 3C points across the whole redshift range
apart from the lowest redshift bin which has a large uncertainty for
the 7C sample due to few objects. We confirm the result of Eales et
al. (1997), Jarvis et al. (2001a) and Inskip et al. (2002) that 6C
radio galaxies are significantly fainter than 3C radio galaxies at
high-redshifts, but appear to be similar at $z<0.6$. Inskip et
al. used a Bayesian statistical approach to show that in fact the two
samples are consistent (at the 95\% level) with having $\Delta K$
between 0.1 and 0.4 magnitudes at all redshifts and so there is no
strong evidence that the apparent difference between the behaviour at
low and high redshifts is significant. Our analysis also confirms this
and shows by the large error bars for the 6C sample at low redshifts
that this uncertainty is due to the small numbers of sources. The 7C
sample, which contains considerably more low-redshift sources, shows
no change in the typical $\Delta K$ at low redshifts. Therefore,
attempts to describe the changing behaviour of the radio-luminosity
dependence of the $K-z$ relation at low redshifts are probably
unnecessary (e.g Best et al. 1998).

In the discussions above we have excluded the possibility that the
correlation of $\Delta K$ with radio luminosity is due to
contamination of the $K$-band flux by non-stellar emission associated
with the AGN, which would be positively correlated to the radio
luminosity. Our data have been corrected for emission line
contamination so this is unlikely to be the cause of the
correlation. Also, Jarvis et al. (2001a) showed that the emission line
contribution to the $K$-band magnitudes is only serious at $z>2$ where
the strong H$\alpha$ and [O\,{\small III}] lines are redshifted into
the $K$-band. Other possible AGN-related emission mechanisms are
partially obscured quasar nuclei, scattered quasar continuum and
jet-induced star-formation. These issues have been investigated in
detail by Leyshon \& Eales (1998), Best et al. (1988), Simpson et
al. (1999). Their conclusions are that the bulk of the $K$-band light
for $z> 1$ 3CRR radio galaxies (which are the most luminous radio
galaxies known) is due to emission from stars. At the highest
redshifts ($z> 3$) it is possible that star formation, maybe induced
by the jets, could provide a dominant contribution to the $K$-band flux (van
Breugel et al. 1998), but this is not the case for most radio galaxies
at $z<2$ (Pentericci et al. 2001). Our finding that the difference
between the $K-z$ relations of 3C and 7C galaxies extends down to the
lowest redshifts also supports the view that AGN-emission is not
responsible for the difference, since at the lowest redshifts even the
3C radio galaxies have only moderately powerful nuclei.

We have shown that there is a significant difference between the $K-z$
relations of 3C and 7C galaxies, however there is not strong evidence
for systematic differences between the 6C and 7C samples in the binned
data. To quantify the amount by which all the samples differ we have
calculated the mean offset from the best fitting $K-z$ relation of
each sample over the redshift range $0.2<z<1.8$. This redshift range
restriction is due to the paucity of 6C and 7C sources at $z<0.2$ and
3C sources at $z>1.8$. The mean $\Delta K$ for each sample is -0.28
for 3C, +0.05 for 6C and +0.27 for 7C. These values are indicated by
the horizontal lines on Fig.\ \ref{fig:kzres} with the grey shading
the $\pm 1 \sigma$ standard errors. They show a systematic trend of
fainter $K$-magnitudes in lower radio luminosity galaxies.

Our finding that the 6C radio galaxies are only 0.3 mag fainter than
those of 3C galaxies is much lower than the 0.6 mag difference found
by Eales et al. (1997). The reason for this different result is that
we have included sources at all redshifts, whereas Eales et al. (1997)
separated the $z>0.6$ and $z<0.6$ galaxies and found the value of 0.6
mag for the high-redshift galaxies only. Our difference of 0.3 mag at
all redshifts fits well with the analysis of Inskip et al. (2002).
McCarthy (1999) showed that the $K-z$ relation for the MRC radio
galaxies (selected at a radio flux-density a factor of four lower than
the 3C sample) follow essentially the same $K-z$ relation as the 3C
sample, with MRC galaxies being only 0-0.2 mag systematically fainter
than 3C galaxies. They claimed this is inconsistent with the 0.6 mag
difference for the 3C and 6C samples from Eales et al. (1997). Apart
from any biases in the MRC sample due to its spectroscopic
incompleteness, it is clear that with our reduced value of the
difference between the 3C and 6C samples, the MRC results are
perfectly consistent.

The 0.55 mag difference in the mean $\Delta K$ for the 3C and 7C
samples, combined with the factor of 20 difference in their median 151
MHz flux-densities, implies that, on average, the stellar luminosity
is higher by a factor $\approx 1.5$ for radio galaxies which are more
powerful by a factor 10.

\subsection{Evolution of host luminosity at constant radio luminosity}

Figs.\ \ref{fig:kzall} and \ref{fig:kzres} have shown that within each
flux-density limited sample the passive evolution description provides a good
fit to the evolution of the $K$-magnitudes. Now we investigate the
apparent cosmic evolution in host luminosities at a constant radio
luminosity. This is important because the rapid decline in the space
density of powerful radio sources at late cosmic epochs (e.g. Dunlop
\& Peacock 1990; Willott et al. 2001) means that powerful radio
sources are triggered much less frequently at low redshifts. But are
they triggered in the same sorts of galaxies?

To investigate this, we use the passive evolution model with $z_{\rm
f}=10$ to determine the $K$-magnitude of an $L_\star$ galaxy as a
function of redshift. A galaxy with this evolution in $K$-magnitude
will therefore have a constant mass with no merging and just passive
evolution of the stellar population. For each galaxy in our sample, we
can now calculate the evolved host luminosity in terms of a multiple
of $L_\star$. The galaxy luminosities range from 0.3 to $14\,L_\star$
with a median of $3.2\,L_\star$. The few highest luminosity objects are
likely to have their $K$-magnitudes brightened by non-stellar emission
or recent star-formation, but these outliers will not strongly affect
our results since only 15 of the 204 sources have luminosities
$>6\,L_\star$.

\begin{figure}
\epsfxsize=0.48\textwidth
\begin{centering}
\epsfbox{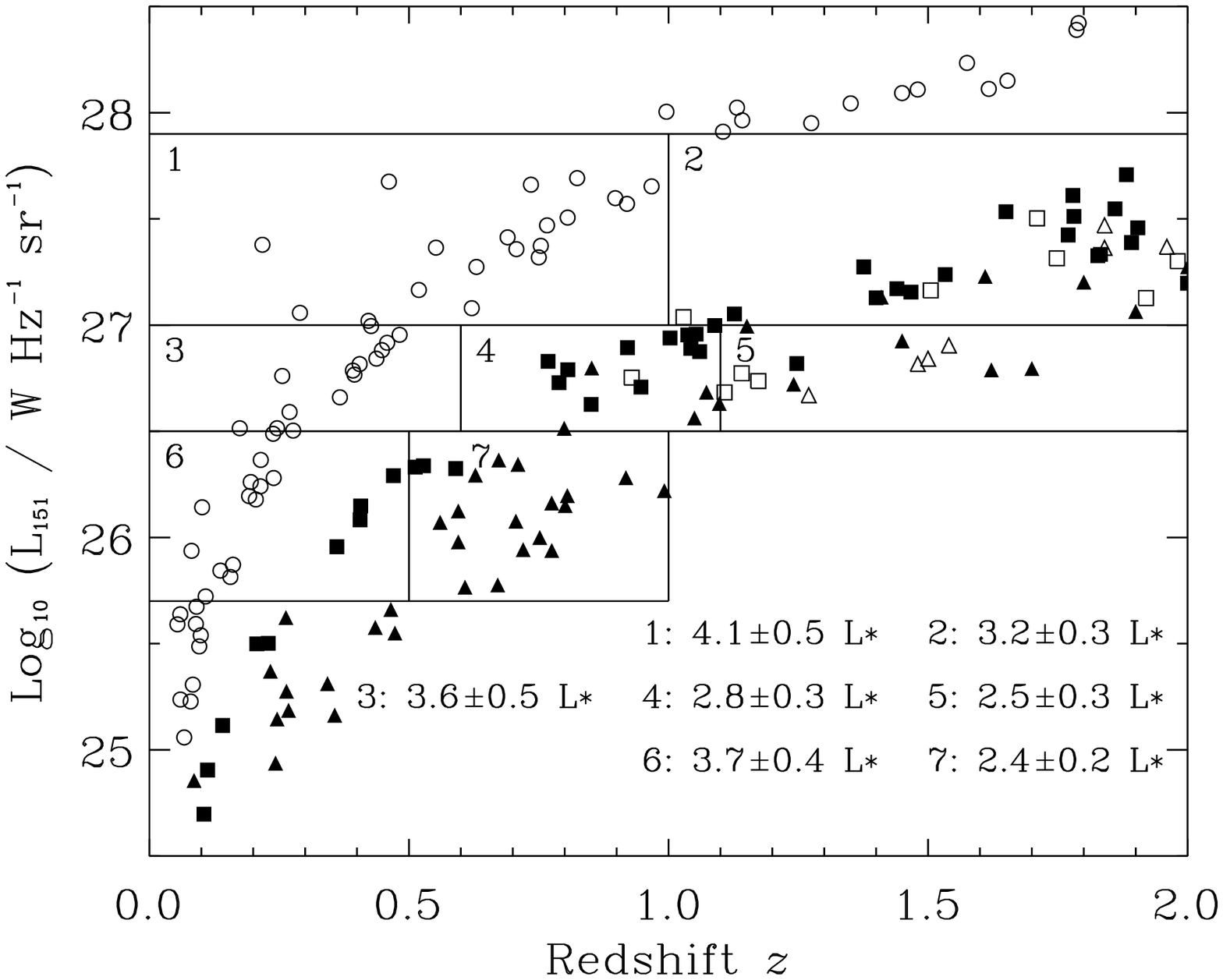} 
\end{centering}

{\caption[junk]{\label{fig:l151} Radio luminosity at 151 MHz
($L_{151}$) against redshift $z$ for radio galaxies in the complete
samples (symbols as for Fig.\ \ref{fig:kzall}) The galaxies have been
binned in terms of $L_{151}$ and $z$ and the mean evolved host
luminosity in each bin is shown in the bottom-right corner of the plot
along with its associated standard error.}}

\end{figure}

We only consider the range in redshift up to $z=2$ because beyond this
the evolution corrections become much more uncertain (e.g the sharp
divergence of the $z_{\rm f}=5$ and $z_{\rm f}=10$ models in Fig.\
\ref{fig:kzall}). In Fig.\ \ref{fig:l151} we plot low-frequency radio
luminosity against redshift for the radio galaxies in the complete
samples up to $z=2$. The galaxies are binned in radio luminosity and
redshift as shown in the figure. The bins were chosen to roughly
separate out the sources into their parent samples and enable the
evolution in host luminosity at a constant radio luminosity to be
determined. Looking at the mean evolved host luminosities shown in the
figure, it is clear that the hosts tend to get relatively less
luminous at higher redshifts. This occurs for all three radio
luminosity slices, although the significance of the change is only
$\sim 2 \sigma$ in each case. The change in host luminosity from low
to high redshifts is $\approx -1\,L_\star$.

The fact that the evolved host luminosities decrease at higher
redshifts can be interpreted as a change in the typical masses of the
galaxies emitting a certain radio luminosity as a function of
redshift. Note that because of our conservative assumption of using
the $z_{\rm f}=10$ passive evolution model, this is a lower limit to
the change in mass with redshift -- any later star-formation than
$z=10$ would tend to increase the change in mass with redshift. This
change in the masses of radio galaxies with redshift can be
interpreted simply with the rapid evolution in the radio luminosity
function from $z=0$ to $z=2$. It is likely that the correlation
between the black hole mass and stellar mass observed at low redshifts
(Magorrian et al. 1998) exists to some extent at all redshifts.  In
this case, the typically higher radio luminosities of galaxies of a
certain mass at higher redshifts may be simply the consequence of a
higher average accretion rate onto a black hole of given mass, perhaps
due to the greater mass of gas available to fuel the activity
(e.g. Archibald et al.\ 2001). Other effects may be important. For
example, the environmental density profiles into which jets of a given
power propagate are likely to be systematically different at different
redshifts. If, because of such environmental effects, sources of a
given radio luminosity have lower average jet power at high z than low
z, then the galaxy mass difference could again be an artifact of a
weak correlation between black hole mass and jet power as discussed in
Sec.\ \ref{lumdep}.

Kauffmann \& Haehnelt (2000) use a semi-analytic $\Lambda$CDM-based
galaxy formation model to predict that the host galaxies of luminous
($M_B \approx -26$) quasars at $z=2$ would be $\approx 5$ times less
massive than at $z=0.5$. The radio galaxies in our highest luminosity
bin have similar radio luminosities to radio-loud quasars with $M_B
\approx -26$. We find the negative evolution in host luminosity (mass)
in our bins with median redshifts $z=0.7$ to $z=1.8$ to be only a
factor of 1.3. Even in a $z_{\rm f}=3$ passive evolution model the
evolution in host mass over this redshift interval would be a factor
of 2.0. This is at odds with the expectations of Kauffmann \& Haehnelt
unless there is a significant difference in the types of galaxies
hosting radio-loud and radio-quiet quasars such as a lower limit to
the mass of radio-loud quasar hosts. However, the similar cosmic
evolution of the two types of quasar (Willott et al. 1998) argues
against such an interpretation. Indeed, Kukula et al. (2001) compared
the host luminosities of radio-quiet quasars with matched nuclear
luminosities at various redshifts out to $z=2$ and found a similar
decrease of at most a factor of two in the host luminosity.

\section{Conclusions}

We have presented $K$-band imaging for the 49 radio galaxies in
regions 7C--I and 7C--II of the 7C Redshift Survey. This sample is
selected at a low radio frequency of 151 MHz and a flux-density level
a factor of 20 lower than the 3CRR sample. The 7C--I and 7C--II data
have been combined with other complete samples to define the
near-infrared Hubble diagram for a total of 205 radio galaxies at
redshifts ranging from 0.05 to 4.4 and investigate the
radio-luminosity dependence of the near-infrared magnitudes.

The $K-z$ relation is well fit by a second order polynomial between
$K$-magnitude and $\log_{10} z$: \begin{equation} K=17.37+4.53
\log_{10} z - 0.31 (\log_{10} z)^{2} \end{equation} This curve is very
close to the expected $K$-magnitude evolution of a passively evolving
instantaneous starburst model galaxy which formed at $z_f=10$ with
present day luminosity of $3\,L_\star$. The rms dispersion about this
relation is found to be 0.58 magnitudes at all redshifts up to
$z=3$. These results are consistent with a high formation redshift
($z_f>5$) for radio galaxies.

Using samples of radio galaxies selected at different flux-density
limits, we have investigated the radio luminosity dependence of the
$K-z$ relation. We find that there is a significant difference between
the $K$-magnitudes of the 3CRR and the fainter 7C radio galaxies over
all redshifts. This is best interpreted as being due to a correlation
of both properties with black hole mass. The typical
evolution-corrected host luminosities decrease at higher redshifts by
a factor in the range 1.3-2. This corresponds to a small decrease in
the masses of radio galaxies at higher redshifts. The weakness of
these correlations of host properties with radio luminosity and
redshift aid the interpretation of the strong cosmic evolution of the
radio source population. The evolution is due to a higher accretion
rate in galaxies of a given mass, presumably due to an increased
available mass of gas as fuel supply.

\section*{Acknowledgements}

We would like to thank Mark Lacy, Steve Eales, Gary Hill and Julia
Riley for their contributions to the 7C Redshift Survey.  We thank the
staff at the UKIRT for their excellent technical support. The United
Kingdom Infrared Telescope is operated by the Joint Astronomy Centre
on behalf of the U.K. Particle Physics and Astronomy Research
Council. We acknowledge the UKIRT Service Programme for some of the
near-infrared imaging. CJW thanks PPARC and the National Research
Council of Canada for support. MJJ acknowledges the support of the
European Community Research and Training Network ``The Physics of the
Intergalactic Medium". KMB thanks the Royal Society for a University
Research Fellowship.

\appendix

\section{Near-infrared images of quasars in 7C--I and 7C--II}

All members of the 7C--I and 7C--II samples have been imaged at
$K$-band by us except the flat-spectrum quasar 5C\,7.230 and the radio
galaxy 3C~200. Redshifts, spectra and multi-wavelength photometry of
the 24 quasars and 2 broad line radio galaxies were presented in
Willott et al. (1998). In this appendix we present the $K$-band images
for these sources so that they can be used as finding charts.

\begin{figure}
\hspace{1.0cm}
\epsfxsize=0.95\textwidth
\begin{centering}
\epsfbox{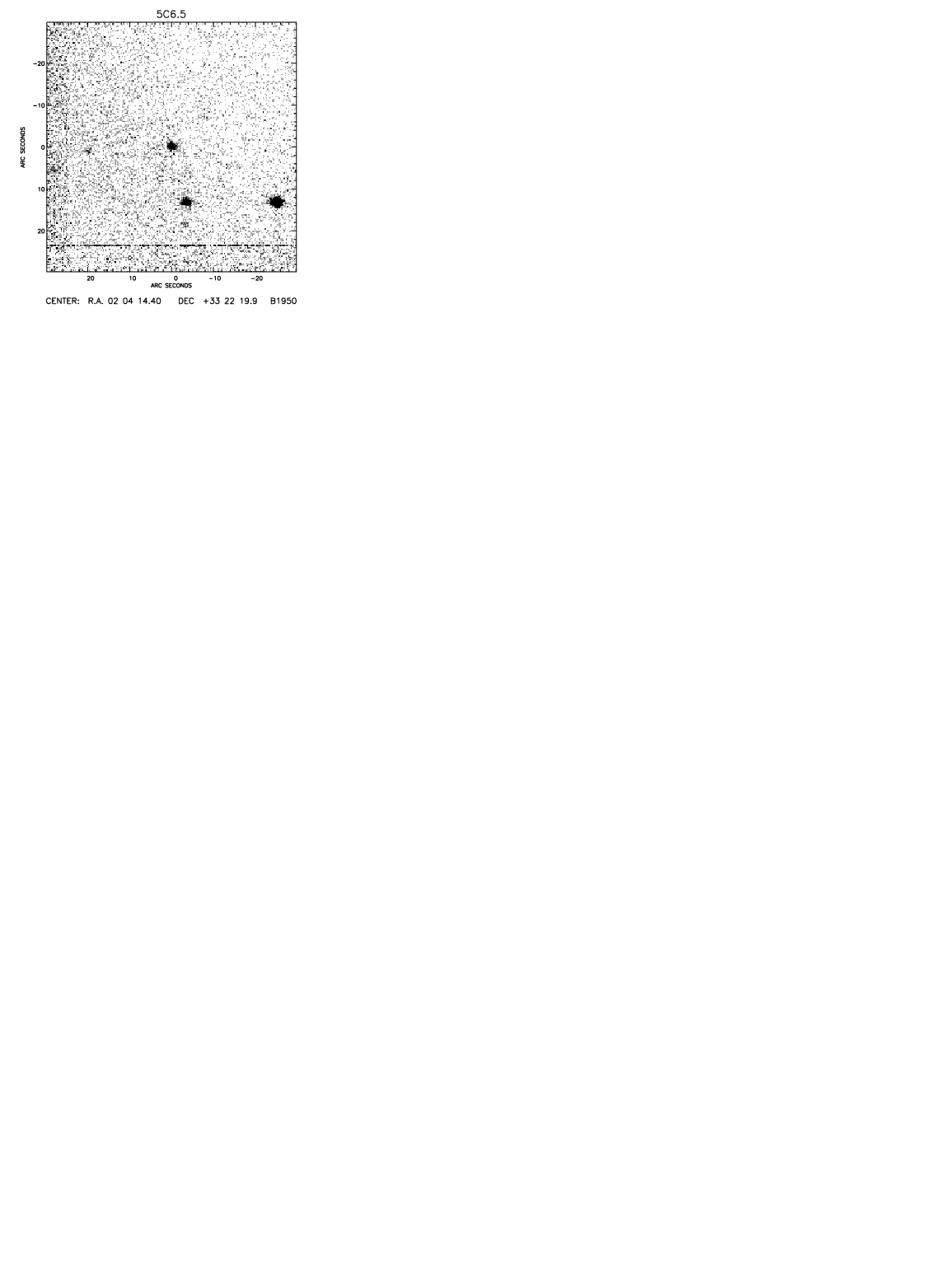} 
\end{centering}
\vspace{-17.7cm}
{\caption[junk]{\label{fig:7ckim6} $K$-band image of a 7C--I quasar. }}
 \end{figure}

\addtocounter{figure}{-1}

\begin{figure*}
\epsfxsize=0.95\textwidth
\begin{centering}
\epsfbox{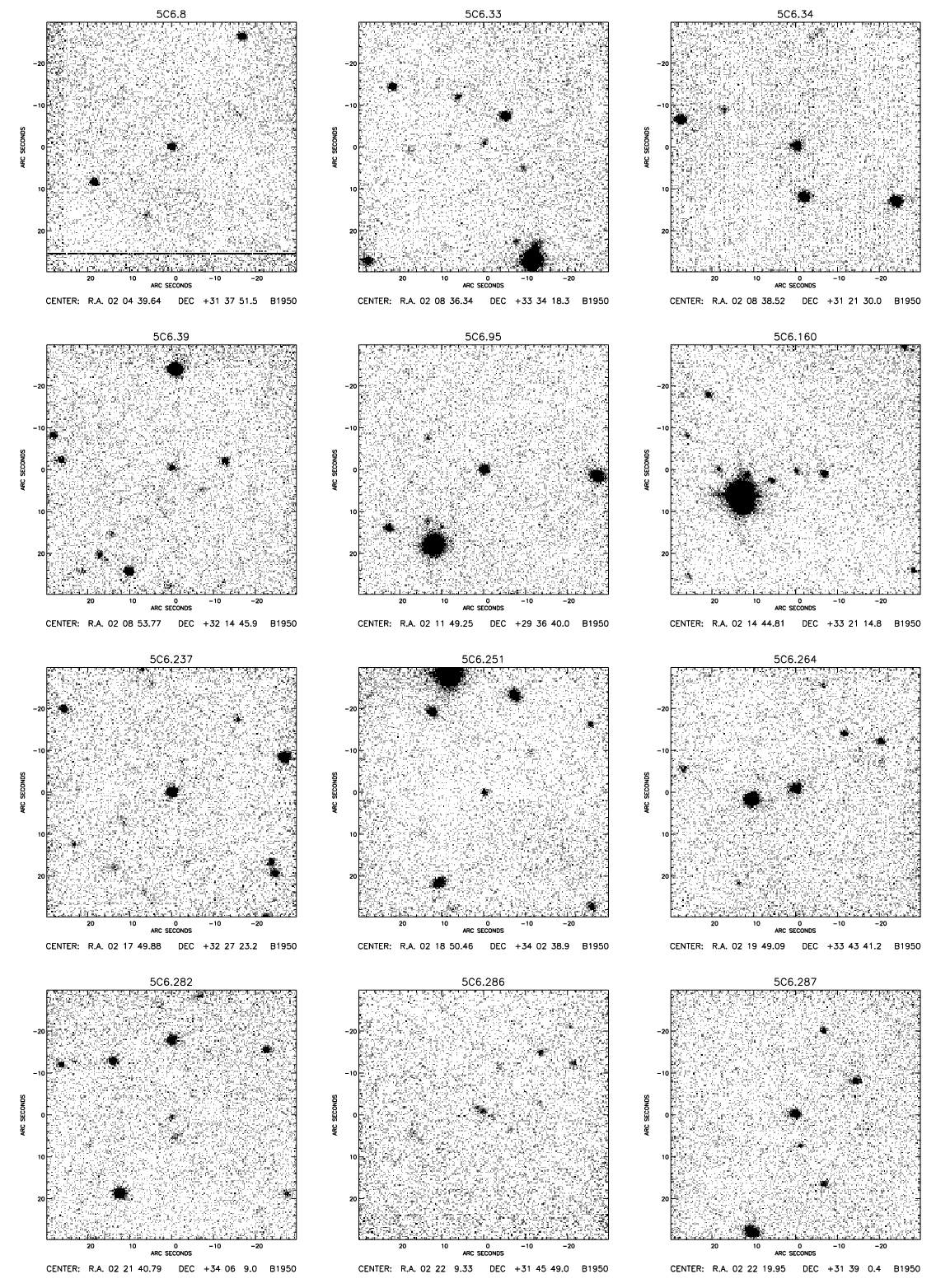} 
\end{centering}
{\caption[junk]{\label{fig:7ckim6}$K$-band images of 7C--I/II quasars.}}
\end{figure*}

\addtocounter{figure}{-1}

\begin{figure*}
\epsfxsize=0.95\textwidth
\begin{centering}
\epsfbox{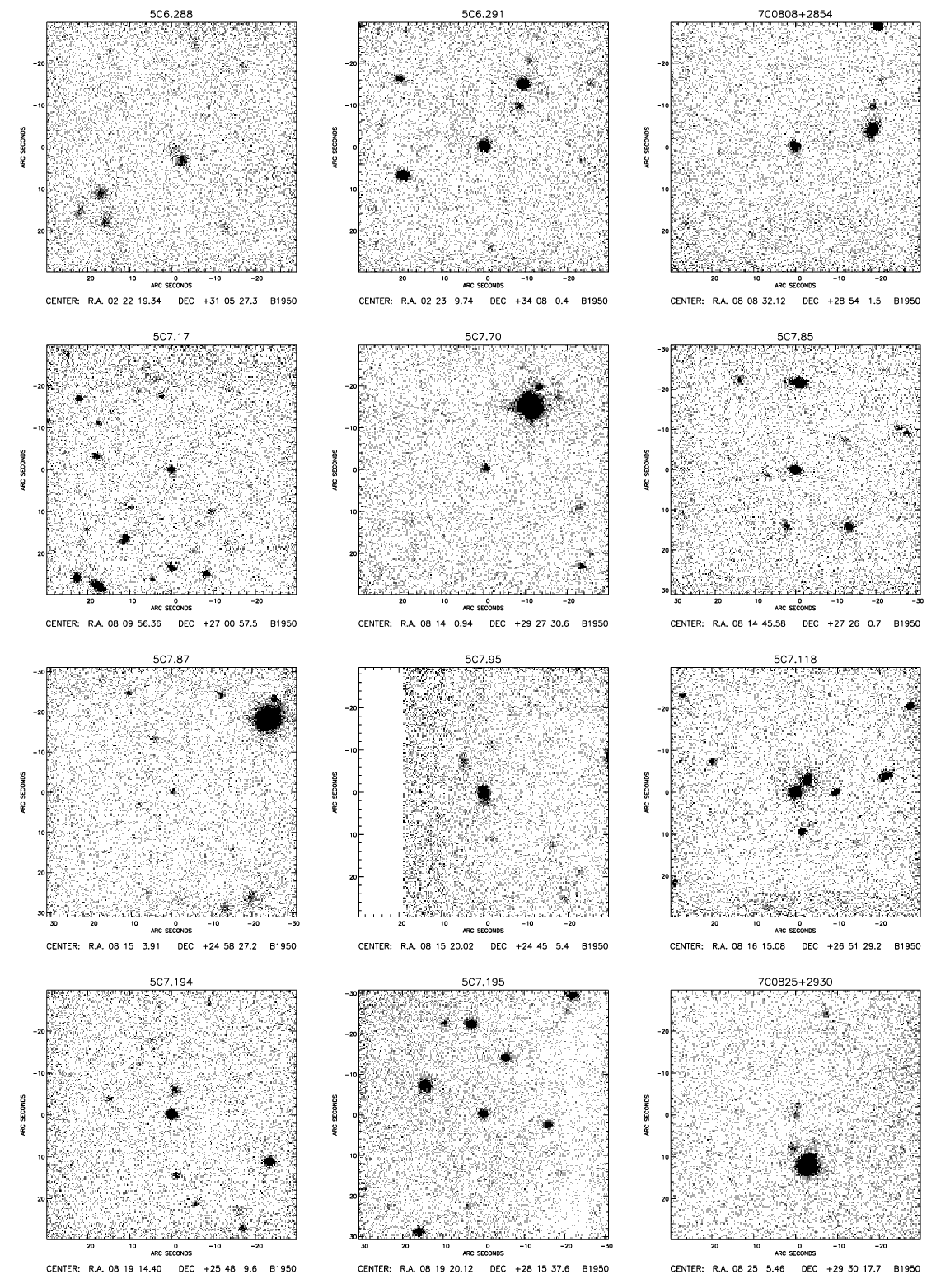} 
\end{centering}
 {\caption[junk]{\label{fig:7ckim6}$K$-band images of 7C--I/II  quasars.}}
\end{figure*}


\begin{thebibliography}{99}

\bibitem{117} Archibald E.N., Dunlop J.S., Hughes D.H., Rawlings S.,
Eales S.A., Ivison R.J., 2001, MNRAS, 323, 417 

\bibitem{160} Best P.N., Longair M.S., R\"ottgering H.J.A., 1998,
MNRAS, 295, 549

\bibitem{170} Bettoni D., Falomo R., Fasano G., Govoni F., Salvo M.,
Scarpa, R., 2001, A\&A, 380, 471

\bibitem{284} Bruzual G., Charlot S., 2002, in prep.

\bibitem{297} Casali M., Hawarden T., 1992, JCMT-UKIRT Newsletter,
No. 4, 33

\bibitem{440} Cole S., Lacey C.G., Baugh C.M., Frenk C.S., 2000,
MNRAS, 319, 168

\bibitem{401} De Breuck C., et al. 2001, AJ, 121, 1241

\bibitem{402} De Breuck C., van Breugel W., Stanford S.A.,
R\"ottgering H.J.A, Miley G., Stern D., 2002, AJ, 123, 637

\bibitem{465} de Vries W.H., O'Dea C.P., Perlman E., Baum S.A.,
Lehnert M.D., Stocke J., Rector T., Elston R., 1998, ApJ, 503, 138

\bibitem{274} Eales S.A., Rawlings S., 1993, ApJ, 411, 67

\bibitem{367} Eales S.A., Rawlings S., Dickinson M., Spinrad H., Hill
G.J., Lacy M., 1993, ApJ, 409, 578

\bibitem{175} Eales S.A., Rawlings S., Law-Green J.D.B., Cotter G.,
Lacy M., 1997, MNRAS, 291, 593

\bibitem{193} Ferrarese L., Merritt D., 2000, ApJ, 539, L9

\bibitem{189} Gebhardt K., et al. 2000, ApJ, 539, L13

\bibitem{164} Hutchings J.B., Neff S.G., 1997, AJ, 113, 550

\bibitem{161} Inskip K.J., Best P.N., Longair M.S., MacKay D.J.C.,
2002, MNRAS, 329, 277

\bibitem{179} Irwin M.J., Maddox S.J., McMahon R.G., 1994, Spectrum:
Newsletter of the Royal Observatories, 2, 14

\bibitem{180} Jarvis M.J., Rawlings S., Eales S.A., Blundell K.M.,
Bunker A.J., Croft S., McLure R.J., Willott C.J., 2001a, MNRAS, 326,
1585

\bibitem{181} Jarvis M.J., et al., 2001b, MNRAS, 326, 1563 

\bibitem{412} Kauffmann G., Haehnelt M., 2000, MNRAS, 311, 576

\bibitem{411} Kukula M.J., Dunlop J.S., McLure R.J., Miller L.,
Percival W.J., Baum S.A., O'Dea C.P., 2001, MNRAS, 326, 1533

\bibitem{309} Laing R.A., Riley J.M., Longair M.S., 1983, MNRAS,
204, 151 

\bibitem{354} Lacy M., Bunker A.J., Ridgway S.E., 2000, AJ, 120, 68

\bibitem{356} Leyshon G., Eales, S.A., 1998, MNRAS, 295, 10 

\bibitem{374} Lilly S.J., Longair M.S., 1984, MNRAS, 211, 833

\bibitem{384} Magorrian J., et al., 1998, AJ, 115, 2285

\bibitem{169} McCarthy P.J., 1999, in `The Most Distant Radio
Galaxies', eds H.J.A. R\"ottgering, P.N. Best and M.D. Lehnert,
Amsterdam, 5

\bibitem{156} Pearson T.J., Kus A.J., 1978, MNRAS, 182, 273

\bibitem{165} Pentericci L., McCarthy P.J., R\"ottgering, H.J.A.,
Miley G.K., van Breugel W.J.M., Fosbury R., 2001, ApJS, 135, 63

\bibitem{267} Rawlings S., Saunders R., 1991, Nature, 349, 138

\bibitem{451} Rawlings S., Lacy M., Leahy J.P., Dunlop J.S.,
Garrington S.T., Ludke E., 1996, MNRAS, 279L, 13

\bibitem{409} Rawlings S., Eales S., Lacy M., 2001, MNRAS, 322, 523

\bibitem{423} Simpson C., Rawlings S., Lacy M., 1999, MNRAS, 306, 828

\bibitem{463} Simpson C., Ward M.J., Wall J.V., 2000, MNRAS, 319, 963

\bibitem{427} Somerville R.S., Primack J.R., 1999, MNRAS, 310, 1087

\bibitem{454} Stockton A., Canalizo G., Ridgway S.E., 1999, ApJ, 519L,
131

\bibitem{268} van Breugel W.J.M., Stanford A.J., Spinrad H., Stern D.,
 Graham J.R, 1998, ApJ, 502, 614

\bibitem{21} Willott C.J., 2001, in `AGN in their Cosmic Environment',
eds. B. Rocca-Volmerange \& H. Sol, EDPS Conf. Series in Astron. \&
Astrophysics, 109

\bibitem{13}  Willott C.J., Rawlings S., Blundell K.M., Lacy M., 1998,
MNRAS, 300, 625

\bibitem{14}  Willott C.J., Rawlings S., Blundell K.M., Lacy M., 1999,
MNRAS, 309, 1017

\bibitem{18} Willott C.J., Rawlings S., Blundell K.M., 2001, MNRAS,
324, 1 

\bibitem{20} Willott C.J., Rawlings S., Blundell K.M., Lacy M., Hill
G.J., Scott S.E., 2002, MNRAS, 335, 1120


\end{thebibliography}
\end{document}